\documentclass[a4paper,10pt,fleqn]{article}
\pdfoutput=1
\usepackage{jheppub}
\usepackage[T1]{fontenc}
\usepackage{diagbox}
\usepackage{setspace}
\usepackage{enumitem}
\usepackage{amsmath}
\usepackage{amsthm}
\usepackage{bm}
\usepackage{mathtools}
\usepackage{braket}
\usepackage{caption}
\usepackage{subcaption}
\usepackage{tikz}
\usepgflibrary{arrows}
\usetikzlibrary{calc}
\usepackage{comment}

\renewcommand{\vec}[1]{\mathbf{#1}}

\renewcommand{\vec}[0]{\boldsymbol}

\newcommand{\gothicn}{\mathfrak{n}}
\newcommand{\beq}{\begin{eqnarray}}
\newcommand{\eeq}{\end{eqnarray}}
\newcommand{\bfd}{{\boldsymbol d}}
\newcommand{\bfv}{{\boldsymbol v}}
\newcommand{\bfnu}{{\boldsymbol\nu}}
\newcommand{\LGP}{\text{LG}(\vec P)}
\newcommand{\eq}{eq.}
\newcommand{\CM}{\mathcal{M}}
\newcommand{\CO}{\mathcal{O}}

\newcommand{\LuuSavage}[0]{Luu:2011ep}
\newcommand{\LtoK}[0]{Hansen:2014eka}
\newcommand{\KtoM}[0]{Hansen:2015zga}
\newcommand{\KSS}[0]{Kim:2005gf}
\newcommand{\LL}[0]{Lellouch:2000pv}
\newcommand{\allInverseLGS}[0]{Beane:2007qr,Hansen:2015zta,Hansen:2016fzj,Muller:2020vtt,NPLQCD:2020ozd}
\newcommand{\twoExtensions}[0]{Rummukainen:1995vs,He:2005ey,Kim:2005gf,Lage:2009zv,Bernard:2010fp,Fu:2011xz,Hansen:2012tf,Briceno:2012yi,Guo:2012hv,Briceno:2014oea}
\newcommand{\three}[0]{Briceno:2012rv,Polejaeva:2012ut,Hansen:2014eka,Hansen:2015zga,Briceno:2017tce,Hammer:2017uqm,Hammer:2017kms,Mai:2017bge,Briceno:2018aml,Briceno:2018mlh,Jackura:2019bmu,Blanton:2019igq,Briceno:2019muc,Romero-Lopez:2019qrt,
Blanton:2020gha,Blanton:2020jnm,Hansen:2020zhy,Blanton:2020gmf}
\newcommand{\writingFaszeta}[0]{Luscher:1986pf,Luscher:1990ux,Rummukainen:1995vs,Kim:2005gf}

\newcommand{\smearedSF}[0]{Hansen:2017mnd,Bulava:2019kbi,Hansen:2019idp,Gambino:2020crt,Bruno:2020kyl}
\newcommand{\fvsg}[0]{Luscher:1990ux,Rummukainen:1995vs,Moore:2005dw,Dudek:2012gj}

\newcommand{\FirstMR}[0]{eq:LOGeneralResult}
\newcommand{\SecondMR}[0]{eq:NNLOCoM}
\newcommand{\ThirdMR}[0]{eq:Delta1AllEllMF}

\title{Analytic expansions of multi-hadron finite-volume energies: I. Two-particle states}

\author[a]{D.~M.~Grabowska}
\author[b]{and M.~T.~Hansen}

\affiliation[a]{Theoretical Physics Department, CERN, 1211 Geneva 23, Switzerland}
\affiliation[b]{Higgs Centre for Theoretical Physics, School of Physics and Astronomy, The University of Edinburgh, Edinburgh EH9 3FD, UK}
\emailAdd{dorota.grabowska@cern.ch}
\emailAdd{maxwell.hansen@ed.ac.uk}

\preprint{CERN-TH-2021-133}

\abstract{We derive analytic expansions for the finite-volume energies of weakly-interacting two-particle systems, using the general relations between scattering amplitudes and energies derived by L{\"u}scher and others. The relations hold for ground and excited states with both zero and non-zero total momentum in the finite-volume frame. A number of instructive aspects arise in the derivation, including the role of accidental degeneracies and the importance of defining a power-counting scheme in the expansions. The results give intuition concerning the imprint of weakly-interacting systems on the energy spectrum, while also providing a useful basis for the analogous results concerning three-particle excited states, to appear.}

\begin{document}
\maketitle
\flushbottom
\abovedisplayskip 11pt
\belowdisplayskip 11pt

\clearpage

\section{Introduction}

An overwhelming body of evidence has established quantum chromodynamics (QCD) as a precise and quantitative description of the strong nuclear force over an incredible range of energies. However, due to a mismatch between the fundamental fields of the theory (quarks and gluons) and the low-energy degrees of freedom (bound states of quarks and gluons, called hadrons), extracting first-principles predictions can be challenging. This is especially true for moderate-energy multi-hadron processes, for which both low-energy effective theory and high-energy perturbative methods break down.

Lattice QCD is a proven method for reliably determining the properties of QCD, especially where analytic techniques fail, by making use of Monte Carlo importance sampling to numerically estimate the quantum path integral, regulated via discretization on a finite spacetime grid. This method has reached an era of sub-percent precision for many single-hadron quantities and has also proven reliable in extracting more complicated, two-hadron observables. Examples in the latter category include two-meson, meson-baryon and baryon-baryon scattering amplitudes as well as one-to-two decay and transition amplitudes. See refs.~\cite{Briceno:2017max,Bulava:2019hpz,Detmold:2019ghl,Edwards:2020rbo,Christ:2020kow,Aoki:2020bew} for recent reviews. More recently, the reach of these calculations has extended to three-to-three scattering amplitudes. See refs.~\cite{Hansen:2019nir,Rusetsky:2019gyk,Mai:2021lwb} for reviews of the progress in this sector.

The standard methodology in the majority of multi-hadron calculations is to use the finite system size of the calculation (the finite volume) as a tool in probing physical observables. In particular, when the fields are constrained to have periodicity $L$ in the three spatial directions, then the continuum of multi-hadron energies is replaced with a discrete set, denoted $E_n(L)$. One can then use field theoretic methods to relate the values of these energies to scattering amplitudes. A modern, rigorous, and general formulation of this idea was provided by L{\"u}scher in refs.~\cite{Luscher:1986pf,Luscher:1990ux}, in which he related the two-to-two elastic scattering amplitude of identical spin-zero particles to finite-volume energies with vanishing total spatial momentum, $\boldsymbol P = \boldsymbol 0$, in the finite-volume frame. This has since been generalized to include non-zero momentum, multiple two-particle channels of non-identical and non-degenerate particles, as well as particles with non-zero spin \cite{\twoExtensions}. Most recently, the methods have been extended to the extraction of amplitudes involving three hadrons in either the initial or final state \cite{\three}.

The purpose of this work, together with a second article to appear, is to derive analytic relations for two- and three-particle finite-volume energies of weakly interacting systems, in the low-energy regime for which only a single channel of identical scalar or pseudo-scalar particles can propagate. The results are directly applicable, for example, to maximal isospin multi-pion and multi-kaon channels is QCD, as well as other weakly-interacting multi-boson systems, including calculations of non-QCD lattice field theories. The present work focuses on two-particle states and extends previous derivations by providing expansions for ground and excited-state energies with any value of total momentum $\boldsymbol P$ in the finite-volume frame as well as describing the contribution of all angular momentum components.

The general relation between energies and scattering amplitudes, restricted to the regime of a single two-particle channel, can be packaged into a single master-function, depending on the energy $E$, the total three-momentum in the finite-volume frame $\boldsymbol P$, and the volume $L$, as well as the scattering amplitude across all partial waves $\ell$, denoted $\mathcal M_{\ell}$, and the specified irreducible representation (irrep) of the finite-volume energies of interest. The latter is defined with respect to the symmetry group of the system, either the full octahedral group including parity or the little group thereof that leaves $\boldsymbol P$ invariant. The function is constructed such that its roots in $E$, with all other inputs held fixed, give the discrete finite-volume spectrum $E_n(L)$ of the system. The condition that this function should vanish is referred to as a quantization condition. To make use of the relations in practice, one must generally truncate the system by approximating $\mathcal M_{\ell} = 0$ for $\ell > \ell_{\sf{max}}$. This work considers both the case of $S$-wave dominance, for which $\ell_{\sf max}=0$, as well as the effects of higher partial waves, detailed in sections \ref{subsec:HPW} and \ref{subsec:AccDeg}. As described in section \ref{subsec:HPW}, it is even possible to write the leading contribution from all partial waves in a compact form. (See also ref.~\cite{Luscher:1986pf}.)

As we discuss extensively in section \ref{subsec:AccDeg}, special care must be taken care for the so-called accidentally degenerate states, already discussed in refs.~\cite{Luscher:1986pf,\LL,\LuuSavage}. To define these, note that a theory of two non-interacting particles in a periodic cubic volume of length $L$, is characterized by assigning a momentum to each, quantized as an integer-vector multiple of $2 \pi/L$. For $\boldsymbol P= \boldsymbol 0$ the two momenta are back-to-back so that a single three-vector characterizes the state. In this case the first accidentally degenerate state is the 8$^{\text{th}}$ excited state, for which non-interacting pions can have back-to-back momentum of type $(0,0,3)$ or else of type $(1,2,2)$. In section \ref{subsec:AccDeg}, we describe such states in detail, for various values of $\boldsymbol P$, and discuss the role higher angular momenta in breaking the degeneracy.

In addition to providing the basis for our subsequent work on three-particle states, we envision a number of broader applications for the results presented here. These include
\begin{itemize}
\item Building general intuition on the interaction-induced shifts to finite-volume energies,
\item Better understanding cancellations leading to exponentially suppressed volume effects in correlators, given the power-like volume dependence of energies in their spectral decompositions,
\item Guiding automated root finders of the full two-particle quantization condition,
\item Exploring the convergence of contributions from higher-partial waves to finite-volume energies.
\end{itemize}
The question of how the value of $E_n(L)$ converges as a function of $\ell_{\sf max}$ has also recently been studied in ref.~\cite{Lee:2021kfn}. We comment in more detail on these items in our conclusion.

We emphasize that the results presented here hold only in the energy regime where the system is weakly interacting and break down, for example, in the vicinity of a narrow resonance. However, even in a resonant system, the expansion will give a good description for any fixed state at sufficiently large $L$. This is because the energy sampled for a given state decreases with increasing $L$, eventually moving away from the resonant behavior and closer to threshold. An exception to this is systems at unitarity, for which the scattering amplitude has a pole at production threshold.

It is important to put this work in context of previously published results concerning expansions of finite-volume energies for weakly-interacting systems. Already in refs.~\cite{Luscher:1986pf,Luscher:1990ux}, L{\"u}scher considered the large-volume expansion of both the ground state and excited states without accidental degeneracy, for $\boldsymbol P= \boldsymbol 0$. The ground-state result matched earlier work of Huang and Yang~\cite{Huang:1957im}, who gave an expression for any number of particles, focusing on the special case of non-relativistic hard spheres. This was later generalized and re-derived in various contexts in refs.~\cite{\allInverseLGS}, to give a general result for the ground-state finite-volume energy of any number of relativistic bosons. More recently, an expansion of three-particle excited states has been discussed in refs.~\cite{Pang:2019dfe,Romero-Lopez:2020rdq}. Finally, in ref.~\cite{\LuuSavage}, Luu and Savage provide an extensive exploration of the phenomenology of L{\"u}scher's quantization condition, including detailed discussion of the finite-volume symmetry group, the role of accidental degeneracies and prospects for extracting higher angular-momentum components from numerical lattice calculations.

This work differs from the results summarized above in three key ways. First we argue that the expansion for excited states is non-unique and can only be performed in the context of a particular power-counting scheme. The latter should be defined to give the best description of the exact energy for a given set of scattering parameters. In particular, we argue that the large-volume expansion is misleading in the sense that $L$ dependence arises in various kinematic factors, including the Lorentz boost factor and the non-interacting energy, order-by-order in the expansions that we perform. Re-expanding these about infinite $L$ is required to recover earlier work but we find this degrades the descriptive power without simplifying the results and is not needed. Second we give a more general framework for covering a wide-range of cases including the contribution of any number of partial waves in the expansion and the general strategy for treating accidentally degenerate states. Third and finally we give results for the ground and excited state expansions for nonzero $\boldsymbol P$ that, as far as we know, have not been considered previously. The final point is of particular importance as it is a crucial input for the corresponding expansions of three-particle energies considered in the companion manuscript, to appear.

The remainder of this work is organized as follows: In the next section we derive the general expansion of two-particle states with any momentum $\boldsymbol P$ in the finite-volume frame, taking particular care to establish a general notation, to specify the role and non-uniqueness of power-counting schemes in defining the expansion, and to explore the effects of higher partial waves, including the case of accidental degeneracy.
The main concrete results of this work are summarized in eqs.~\eqref{\FirstMR}, \eqref{\SecondMR}, \eqref{\ThirdMR}, in tables~\ref{tab:Pell}-\ref{tab:Pell002}, and in the discussion of section~\ref{subsec:AccDeg}.
Then, in section \ref{sec:numerical}, we describe various numerical checks performed by comparing our expansion to the general solutions of the quantization condition.
In section \ref{sec:Conc} we briefly conclude. This work additionally contains five appendices, addressing more technical aspects of the derivation including a description of the finite-volume functions required for the quantization condition, a comment on energy level crossings, a summary of the relevant symmetry groups and projectors, and an important relation on non-interacting energies required to perform the expansion.

\section{Derivation and results}\label{sec:Derivation}

The following subsections review the general formalism and establish the expansion strategy and the notation used. The main results are emphasized at the end of each subsection.

\subsection{Set-up}

Our aim is to perturbatively expand finite-volume two-particle energies, collectively denoted $E_{n, \vec P, \Lambda}(L)$, about the non-interacting limit. Here the integer index $n$ indicates the $n^{\text{th}}$ excited state ($n=0$ the ground state), $\vec P$ is the total spatial momentum of the two-particle system, $\Lambda$ is the relevant finite-volume irrep, and $L$ is the box length. The total momentum $\vec P$ is equal to an integer-vector multiple of $(2 \pi/L)$. We write this as $\vec P = (2 \pi/L) \vec d = (2 \pi/L)(d_x,d_y,d_z)$ and use the shorthand $\vec P = [d_xd_yd_z]$. In this work we restrict attention to a single channel of identical scalar particles with mass $m$.

The simplest example of the expressions derived in this work is that of the two-particle ground state $(n=0)$ for vanishing spatial momenta $\vec P=[000]$ in the trivial-irrep $\Lambda = A_{1g}$ of the octahedral group $O_h$. The result, through $\mathcal O(1/L^5)$, was presented by Huang and Yang in ref.~\cite{Huang:1957im} and subsequently extended to higher orders.
In our notation the first non-trivial order reads
\begin{equation}
E_{0,[000],A_{1g}}(L) = 2 m + \frac{4 \pi a_0}{m L^3} + \mathcal O(1/L^4) \,,
\label{eq:SWaveLO}
\end{equation}
where $a_0$ is the two-particle scattering length, defined in \eq~\eqref{eq:EffRangeExp} below.

The key tool used to derive such results here is the finite-volume quantization condition \cite{Luscher:1990ux,\KSS,Rummukainen:1995vs}
\beq
\label{eq:QC}
\det_{\Lambda \mu} \! \Big [ \mathbb P_{\Lambda, \mu} \cdot \big [ \CM(E^\star)^{-1} + F(E, \vec P, L) \big ] \cdot \mathbb P_{\Lambda, \mu} \Big ] \bigg \vert_{E \, = \, E_{n,\vec P, \Lambda}(L)} = 0 \,,
\eeq
where $\CM(E^\star)$ is the infinite-volume scattering amplitude and $F$ is a matrix of known geometric functions, reviewed in appendix~\ref{app:FFunction}. The matrix $F$ depends on the total energy and momentum in the finite-volume frame, $(E, \boldsymbol P)$, as well as the box length, $L$. By contrast, the scattering amplitude only depends on the center-of-momentum frame (CoM frame) energy
\begin{equation}
\label{eq:CoMEnergy}
E^\star = \sqrt{E^2 - \boldsymbol P^2} \,.
\end{equation}
We have also introduced $\mathbb P_{\Lambda, \mu}$ as a projector restricting to the irrep of interest. Some discussion of finite-volume groups and projectors is included in appendix \ref{app:fvsg}. See also refs.~\cite{\fvsg} for more details.

The matrices $\mathcal M(E^\star)$, $F(E, \boldsymbol P, L)$ and $\mathbb P_{\Lambda, \mu}$ are each defined on an angular momentum space and carry two sets of spherical harmonic indices, e.g.~$\mathcal M(E^\star) = \mathcal M_{\ell' m', \ell m}(E^\star)$ where $\ell = 0, 1, 2 \cdots$ and $m = - \ell, - \ell+1, \cdots \ell$. The explicit definition can be given in terms of the $\ell^{\text{th}}$ scattering phase shift, $\delta_{\ell}$. In the case of a single channel of identical scalar (or pseudoscalar) particles, one has
\beq
\CM_{\ell' m', \ell m}(E^\star) = \delta_{\ell' \ell}\delta_{m' m} \frac{16 \pi E^\star}{p^\star}\frac{ e^{2 i \delta_{\ell}(p^\star)}-1}{2i} = \delta_{\ell' \ell}\delta_{m' m} \frac{ 16 \pi E^\star}{p^\star \cot \delta_{\ell}(p^\star) - i p^\star} \,,
\label{eq.ScattAmp}
\eeq
and scattering length, $a_0$, appearing in eq.~\eqref{eq:SWaveLO} is defined via the leading order expansion of $p^\star \cot \delta_{0}(p^\star)$ about threshold
\begin{equation}
p^\star \cot \delta_{0}(p^\star)= - \frac{1}{a_0} + \mathcal O(p^{\star 2}) \,.
\label{eq:pCdThreshold}
\end{equation}
Here we have also introduced the relative momentum
\beq
p^{\star 2} = \frac{E^{\star 2}}{4} - m^2 \,.
\label{eq: EnergyMomentumRelation}
\eeq
See table~\ref{tab:kin} for a summary of the notation established so far, and continued in the following. 

\renewcommand{\arraystretch}{1.5}
\begin{table}
\begin{center}
\begin{tabular}{c | c | l}
\ \ Quantity \ \ & \ \ Definition/Key relation \ \ & \ \ \ Description \\ \hline \hline
$L$ & --- & \ \ \ finite-volume box length \\
$E$ & --- & \ \ \ finite-volume-frame energy \\
$n$ & --- & \ \ \ level of the excited state \\
$\vec P$ & $2 \pi \vec d /L$ & \ \ \ finite-volume-frame 3-momentum \\
$E^\star$ & $ \sqrt{E^2 - \vec P^2}$ & \ \ \ center-of-momentum-frame energy \\
$\vec k$ & $ 2 \pi \vec v/L$ & \ \ \ generic 3-momentum\\
$\omega_k$ & $\sqrt{\vec k^2 + m^2}$ & \ \ \ on-shell time component of 4-vector $k^\mu$ \\
$\ell m$ & indices on $Y_{\ell m}$ & \ \ \ angular-momentum indices \\
$p^\star$ & $\sqrt{E^{\star2}/4 - m^2}$ & \ \ \ relative momentum magnitude from $E^\star$ \\
$q $ & $L p^\star/(2 \pi)$ & \ \ \ dimensionless version of $p^\star$\\ \hline
$\mathfrak{n} $ & $\{n, \vec P, \Lambda\}$ & \ \ \ collective index for a state $\mathfrak{n} $ \\
$ E_\mathfrak{n}^{(0)}(L)$ & eq.~\eqref{eq:E0ndef} & \ \ \ non-interacting energy \\
$\boldsymbol \nu_\mathfrak{n} $ & --- & \ \ \ representative integer three-vector for state $\mathfrak{n} $ \\
$q_\mathfrak{n}^{(0)2}$ & $L^2 [E_\mathfrak{n}^{(0)}(L) - \boldsymbol P^2 - 4 m^2 ]/(2 \pi)^2$ & \ \ \ non-interacting dimensionless momentum \\[4pt]
$\boldsymbol S_\mathfrak{n}$ &
eq.~\eqref{eq:E0ndefForn}
& \ \ \ set of dimensionless momenta \\[-8pt]
&& \ \ \ that give same non-interacting energy \\[-2pt]
\end{tabular}
\caption{Summary of the notation used throughout the paper. Note that $q$ is only used to denote the dimensionless momentum in the center-of-momentum (CoM) frame. We have dropped the $\star$, which generically indicates CoM quantities, for simplicity in this case.\label{tab:kin}}
\end{center}
\end{table}
\renewcommand{\arraystretch}{1.0}

As the matrices are formally infinite dimensional, in practice one must truncate the quantization condition by setting $\mathcal M_{\ell' m', \ell m}(E^\star) = 0$ or equivalently $\delta_{\ell}(p^\star) = 0$ for $\ell > \ell_{\sf{max}}$. As discussed in refs.~\cite{Luscher:1986pf,Luscher:1990ux,Kim:2005gf}, one can then set the corresponding entries of $F$ and $\mathbb P_{\Lambda, \mu}$ to zero without further approximation. Given a truncated version of \eq~\eqref{eq:QC}, the expansion is derived by substituting
\begin{equation}
E_\mathfrak{n}(L) = E_\mathfrak{n}^{(0)}(L) + \sum_{k=1}^\infty \epsilon^k \, \Delta^{(k)}_{E[\mathfrak{n}]}(L) \,,
\label{eq:EnergyExpansion}
\end{equation}
where $\mathfrak{n} = \{n, \vec P, \Lambda\}$ is a collective index for all discrete information common to each building block and $E_\mathfrak{n}^{(0)}(L)$ is the finite-volume energy in the non-interacting limit, given e.g.~by taking $\delta_{\ell} \to 0$ for all $\ell$. This non-interacting energy can be written as
\begin{equation}
\label{eq:E0ndef}
E_\mathfrak{n}^{(0)}(L) = \sqrt{m^2 + (2 \pi/L)^2 \vec \nu_{\mathfrak{n}}^2} + \sqrt{m^2 + (2 \pi/L)^2 (\vec d -\vec \nu_{\mathfrak{n}} )^2} \,,
\end{equation}
where $\vec \nu_{\mathfrak{n}}$ is an integer vector representing the non-interacting state. For most values of ${\mathfrak{n}}$, multiple choices of $\vec \nu_{\mathfrak{n}}$ are possible, so that this identifier is not unique. All quantities that depend on this vector, e.g.~the non-interacting energy, $ E_\mathfrak{n}^{(0)}(L)$, are equal for any valid choice.
The correspondence between $n$ and $\boldsymbol \nu_{\{n,\boldsymbol P, \Lambda\}}$ is defined such that $ E_\mathfrak{n}^{(0)}(L) < E_{\mathfrak{n}'}^{(0)}(L)$ for $n<n'$ \emph{in the large $L$ limit}. As we prove in appendix~\ref{app:NoLevelCrossing}, the sorting is independent of $L$ in the CoM frame, but non-interacting level crossings can occur for non-zero total momenta.

Combining eqs.~\eqref{eq:QC} and \eqref{eq:EnergyExpansion},
it is possible to solve for the corrections $\Delta_{E[\mathfrak n]}^{(k)}$ order by order in terms of $\epsilon$. As we explain in section~\ref{subsec:PowerCS} below, to completely define the expansion one must assign an epsilon scaling to various parameters entering $\mathcal M(E^\star)$ as well as, possibly (but not neccesarily), to $1/L$. There is no unique assignment and the best choice of power-counting scheme depends on the details of the system.

\subsection{$S$-wave dominance: Leading-order shift}\label{sec:sWaveOnly}

We begin with the expansion for the case of $\ell_{\sf{max}} = 0$. For this truncation, only the trivial irreps ($A_{1g}$ for $\vec P = [000]$ and $A_1$ otherwise) have finite-volume energies that are shifted by interactions. These are found by solving
\begin{align}
p^\star \cot \delta_0(p^\star) & = f(q,\boldsymbol d, L) \,,
\label{eq:SWaveMaster}
\end{align}
where
\begin{equation}
\label{eq:fdef}
f(q,\boldsymbol d, L) \equiv - 16 \pi E^\star \, \text{Re} \big [F_{00,00}(E, \vec P, L) \big ]
= - \lim_{s \to -1} \frac{1}{\gamma(q,\boldsymbol d, L) \, \pi L}\sum_{\vec v \in \mathbb Z^3} \Big [ q^{2} - \Gamma(\vec v \vert q, \vec d, L) \Big ]^{\! s} \,.
\end{equation}
The second equality is a standard definition of $F_{00,00}$, first introduced for nonzero $\vec d$ in ref.~\cite{Rummukainen:1995vs}, and the limit indicates that the sum is regulated by analytically continuing from $\text{Re}[s]<-1$ \cite{Luscher:1990ux}. The relation between this and other standard definitions, and the corresponding expressions for general $F_{\ell' m',\ell m}(E, \boldsymbol P, L)$, are reviewed in appendix \ref{app:FFunction}.
In eq.~\eqref{eq:fdef} we have also adjusted the coordinate dependence, changing from $E, \boldsymbol P, L$ to $q,\boldsymbol d, L$, with the latter including
\beq
q^2 = \left(\frac{L}{2\pi}\right)^{\!2}\left(\frac{E^2}{4} - \frac{\boldsymbol P^2}{4} - m^2 \right) \,, \qquad \qquad \gamma(q,\boldsymbol d, L) = \bigg [ 1 + \frac{ \boldsymbol d^2}{4 (q^2 + [m L/(2 \pi)]^2 ) } \bigg ]^{1/2} \, .
\label{eq:qDefgDef}
\eeq
As is summarized in table~\ref{tab:kin}, $q$ is proportional to $p^\star$, made dimensionless by $L/(2\pi)$. Although we generically use a $\star$ superscript to denote CoM frame quantities, we drop this for $q$ to avoid over-cluttering the notation in the following.
The quantity $f(q,\boldsymbol d, L)$ also depends on
\beq
\Gamma(\vec v \vert q, \vec d, L)\equiv \frac{1}{\gamma(q, \vec d, L)^2}\left(\! \frac{\vec v \cdot \vec d}{\vert \boldsymbol d \vert} - \frac{\vert \boldsymbol d \vert}{2}\right)^2 + \bigg( \frac{\vec v \cdot \vec d}{ \vert \boldsymbol d \vert^2} \, \boldsymbol d - \boldsymbol v \bigg )^2 \,,
\eeq
denoted by $r^2$ in ref.~\cite{Rummukainen:1995vs}.

Since both the left- and the right-hand sides of eq.~\eqref{eq:SWaveMaster} are analytic functions of $q^2$, it is easiest to use this quantity in the expansion. We therefore define $q_{\frak n}(L)^2$ by evaluating eq.~\eqref{eq:qDefgDef} at $E_{\frak n}(L)$
and also introduce the analog of eq.~\eqref{eq:EnergyExpansion} above
\begin{equation}
q_\mathfrak{n}(L)^2 = q_\mathfrak{n}^{(0)}(L)^2 + \sum_{k=1}^\infty \epsilon^k \, \Delta^{(k)}_{q[\mathfrak{n}]}(L) \,.
\end{equation}
Here $q_\mathfrak{n}^{(0)}(L)^2$ is the non-interacting version of $q_\mathfrak{n}(L)^2 $, defined by evaluating eq.~\eqref{eq:qDefgDef} at $E^{(0)}_{\frak n}(L)$.

Having introduced all notation, the general procedure is relatively straightforward: Evaluate both sides of eq.~\eqref{eq:SWaveMaster} at $q_\mathfrak{n}(L)^2$ and solve the equation order-by-order $\epsilon$ to determine $ \Delta^{(k)}_{q[\mathfrak{n}]}(L)$. If desired, the result can then be readily converted back to $E_\mathfrak{n}(L)$, which can be re-expanded to fixed order in $\epsilon$.

To begin this iterative procedure, we take $p^\star \cot \delta_0(p^\star) = \mathcal O(1/\epsilon)$ and infer that the leading-order constraint arising from eq.~\eqref{eq:SWaveMaster} is that $f( q_\mathfrak{n} , \boldsymbol d , L )$ must also scale as $1/\epsilon$. This is achieved by requiring
\beq
q_\mathfrak{n}^{(0)}(L)^2 - \Gamma(\vec v \vert q_{\frak n}^{(0)}, \vec d, L) = 0 \,.
\label{eq:MFPoleequation}
\eeq
As was first shown in refs.~\cite{Rummukainen:1995vs,Kim:2005gf} and as we review in appendix \ref{app:DegenVecSolu}
this is consistent with the relation between $q_\mathfrak{n}^{(0)}(L)^2$ and $E^{(0)}_{\frak{n}}(L)$ and with eq.~\eqref{eq:E0ndef}.

To expand beyond the trivial order, we define ${\boldsymbol S}_{\frak n}$ as the set of all integer three-vectors $\vec v$ satisfying eq.~\eqref{eq:MFPoleequation}, equivalently
\beq
\label{eq:E0ndefForn}
{\boldsymbol S}_{\frak n} = \Big \{ \ \boldsymbol v \in \mathbb Z^3 \ \Big \vert \ E_\mathfrak{n}^{(0)}(L) = \sqrt{m+ (2\pi/L)^2 \boldsymbol v^2}+\sqrt{m+ (2\pi/L)^2 (\boldsymbol d - \boldsymbol v)^2} \ \Big \} \,.
\eeq
In general, ${\boldsymbol S}_{\frak n}$ is given by all rotations of both $\boldsymbol \nu_{\frak n}$ and $\boldsymbol d - \boldsymbol \nu_{\frak n}$ by elements of the little group of $\boldsymbol P$, $\LGP$. The exception to this is the case of accidental degeneracies, for which the definition also contains $\boldsymbol v$ that are not related by such a transformation. Instructive examples of $\vec S_{\frak n}$ are collected in table \ref{tab:SetVectors}.

\begin{table}
\begin{center}
\begin{tabular}{c | c | c | c | c }
\ $n$ \ & \ \ $\vec P$ \ \ & \ $\Lambda$ \ & \ \ $\vec S_\mathfrak{n}$ \ \ & \ $g_\mathfrak{n}$ \ \\ \hline \hline
0 & $[000]$ & \ $A_{1g}$ \ & $(0,0,0)$ & 1 \\ \hline
1 & $[000]$ & \ $A_{1g}$ \ & $(0, 0, 1) \ + $ rotations and flips & 6 \\
\hline
8 & $[000]$ & \ $A_{1g}$ \ & $(1,2,2) \ + $ rotations and flips (24 total) & 30 \\
&&& $(0, 0, 3) \ + $ rotations and flips (6 total) & \\ \hline
0 & $[001]$ & \ $A_1$ \ & $(0,0,0), \ (0, 0, 1)$ & 2 \\ \hline
1 & $[011]$ & \ $A_1$ \ & $(0,0,1),\ (0, 1, 0)$ & 2 \\ \hline
0 & $[002]$ & \ $A_1$ \ &$(0,0,1)$& 1 \\ \hline
\end{tabular}
\caption{Examples for the set $\vec S_\mathfrak{n}$ with total number of elements $g_\mathfrak{n}$. In the accidentally degenerate case, $n=8$ and $\vec P =[000]$, all three-vectors satisfying eq.~\eqref{eq:E0ndefForn} are included in the definition.\label{tab:SetVectors}}
\end{center}
\end{table}

Finally, to give the leading-order energy shift, one must make a specific choice for the expansion of $p^\star \cot \delta_0(p^\star)$.
For example, if the power-counting is such that
\beq
p^\star \cot \delta_0(p^\star) = -\frac{1}{a_0}+ \CO(\epsilon^0) \,,
\eeq
formally holds at all $p^\star$, then the next order is solved by identifying the $1/\epsilon$ term within $f(q_{\frak n}, \boldsymbol d, L)$
\begin{equation}
f(q_{\frak n}, \boldsymbol d, L) = - \frac{1}{\epsilon \Delta_{q[\mathfrak{n}]}^{(1)}(L)} \frac{1}{\gamma_\mathfrak{n}^{(0)}}\frac{1}{\pi L} \sum_{\vec v \in \vec S_\mathfrak{n}} \frac{1}{1- \partial_{q^2} \Gamma(\vec v \vert q, \vec d, L) } \bigg \vert_{q_\mathfrak{n}^{(0)}} + \CO(\epsilon^{0}) \,.
\label{eq:fFullLo}
\end{equation}

One can further show that
\beq
\partial_{q^2} \Gamma(\vec v \vert q, \vec d, L) \bigg \vert_{q_\mathfrak{n}^{(0)}, \boldsymbol v \in \vec S_{\frak n}}
=
\bigg ( \frac{ \omega_{\bfnu_\gothicn}-\omega_{\vec d -\bfnu_\gothicn} }{ \omega_{\bfnu_\gothicn} + \omega_{\vec d -\bfnu_\gothicn} } \bigg )^{\!\!2} \,,
\eeq
where we have introduced
\begin{equation}
\omega_{\bfnu_\gothicn} = \sqrt{m^2 + (2\pi/L)^2 \boldsymbol \nu_{\frak n}^2 } \, \,, \qquad \qquad \omega_{\vec d -\bfnu_\gothicn} = \sqrt{m^2 + (2\pi/L)^2 ( \boldsymbol d - \boldsymbol \nu_{\frak n})^2 } \,.
\end{equation}
Simplifying eq.~\eqref{eq:fFullLo}
and using also that $ \Gamma(\vec v \vert q, \vec d, L) $ is the same for all elements of $\vec S_\mathfrak{n}$, we reach
\begin{equation}
\Delta_{q[\mathfrak{n}]}^{(1)}(L) = a_0 \frac{1}{\gamma_\mathfrak{n}^{(0)}}\frac{1}{\pi L} \frac{g_{\frak n} E^{(0)}_\gothicn (L)^2 }{ 4 \omega_{\bfnu_\gothicn} \omega_{\vec d-\bfnu_\gothicn} } \,.
\end{equation}
Here we have taken $\gamma_\mathfrak{n}^{(0)}$ to denote the boost factor evaluated at the $\frak{n}$th non-interacting energy and have introduced $g_{\frak n} $ as the number of elements within $\vec S_\mathfrak{n}$ ($g_{\frak n} = | \vec S_\mathfrak{n} |$).
The corresponding energy can then be written
\beq
E_\gothicn(L) = E^{(0)}_\gothicn (L) + g_{\mathfrak{n}} \frac{E^{(0)}_\gothicn (L) }{ 4 \omega_{\bfnu_\gothicn} \omega_{\vec d-\bfnu_\gothicn} }\frac{8 \pi a_0}{ \gamma_\mathfrak{n}^{(0)} L^3} + \CO(\epsilon^2) \,.
\label{eq:LOGeneralResult}
\eeq

This is the main result of this subsection. The appearance of $4 \omega_{\bfnu_\gothicn} \omega_{\vec d-\bfnu_\gothicn}$ can be understood as a relative normalization factor, arising from the definition of the scattering amplitude in terms of relativistically normalized states. The factor $ g_{\mathfrak{n}}$, which can also be traced to normalization of states, implies that higher-multiplicity finite-volume energies may offer more sensitivity to scattering information, at least for weakly interacting systems. Finally, note that the boost factor multiplies the $L^3$ factor, so that the moving state effectively sees a larger box. This can be roughly interpreted by identifying the $L^3$ periodicity as the geometry resulting after a length contraction such that the underlying rest-frame volume is effectively larger. We stress again that this result only applies for states that are not accidentally degenerate, i.e.~for states in which all elements of $ \vec S_\mathfrak{n} $ are related by transformations of the octahedral group or the relevant moving-frame little group, $\LGP$.

\subsection{$S$-wave dominance: All orders}\label{sec:sWaveOnlyAllOrders}

The approach of the previous subsection can now be readily be generalized to all orders. For the left-hand side of eq.~\eqref{eq:SWaveMaster} one substitutes
\begin{align}
p^\star \cot \delta_0(p^\star) & = \sum_{m=0}^\infty \frac{1}{m!} \mathcal K_m \left(\frac{2\pi}{L}\right)^{\!\! 2m} \big [q^2 - q_\mathfrak{n}^{(0)}(L)^2 \big ]^m \,, \\
& = \sum_{m=0}^\infty \frac{1}{m!} \mathcal K_m \left(\frac{2\pi}{L}\right)^{\!\! 2m} \bigg [ \sum_{k=1}^\infty \epsilon^k \, \Delta^{(k)}_{q[\mathfrak{n}]}(L) \bigg ]^m \,,
\label{eq:PhaseExp}
\end{align}
where we have introduced the coefficients
\begin{equation}
\mathcal K_m \equiv \left(\frac{\partial}{\partial p^{\star 2}}\right)^{\!\!m} p^\star \cot \delta_0(p^\star) \bigg \vert_{p^{\star 2} = [E^{(0)2}_{\frak n} - \boldsymbol P^2]/4 - m^2 } \,.
\label{eq:KFuncDef}
\end{equation}

This is then matched to the expansion of $f(q_{\frak n}, \boldsymbol d, L)$. To give the latter, we first write $f(q_{\frak n}, \boldsymbol d, L)$ in terms of $\gamma$ and two additional functions
\begin{equation}
f(q_{\frak n}, \boldsymbol d, L) \equiv \frac{1}{ \pi L} \frac{\tau_{\frak n}(q_{\frak n}, \vec d, L) + \beta_{\frak n}(q_{\frak n}, \vec d, L) }{\gamma(q_{\frak n}, \vec d, L)} \,,
\end{equation}
where
\begin{align}
\tau_{\frak n}(q, \vec d, L) & \equiv - \sum_{\vec v \in \vec S_\mathfrak{n}} \Big [ q^{2} - \Gamma(\vec v \vert q, \vec d, L) \Big ]^{-1} \,, \\
\beta_{\frak n}(q, \vec d, L) & \equiv - \lim_{s \to -1} \sum_{\vec v \notin \vec S_\mathfrak{n}} \Big [ q^{2} - \Gamma(\vec v \vert q, \vec d, L) \Big ]^{\! s} \,.
\end{align}
Each of these are then expanded in powers of $q^2_{\frak n}(L) - q^{(0)}_{\frak n}(L)^2$. For example, we introduce the coefficients $G_{\frak n ,m}$ and $B_{\frak n, m}$ via
\begin{align}
\frac{1}{\gamma(q_{\frak n}, \vec d, L)}& = \sum_{m = 0}^\infty \frac{1}{m!} G_{\frak n, m}(\vec d, L) \bigg [ \sum_{k=1}^\infty \epsilon^k \, \Delta^{(k)}_{q[\mathfrak{n}]}(L) \bigg ]^m \,, \label{eq:boost} \\
\beta_{\frak n}(q, \vec d, L) & = \sum_{m = 0}^\infty \frac{1}{m!} B_{\frak n, m}(\vec d, L) \bigg [ \sum_{k=1}^\infty \epsilon^k \, \Delta^{(k)}_{q[\mathfrak{n}]}(L) \bigg ]^m \,. \label{eq:beta}
\end{align}
The $\tau$ function, by contrast, starts with a term proportional to the inverse energy shift and can be written as
\begin{equation}
\tau_{\frak n}(q_{\frak n}, \vec d, L) = T_{\frak n, -1}(\vec d, L) \bigg [ \sum_{k=1}^\infty \epsilon^k \, \Delta^{(k)}_{q[\mathfrak{n}]}(L) \bigg ]^{-1} + \sum_{m = 0}^\infty \frac{1}{m!} T_{\frak n,m}(\vec d, L) \bigg [ \sum_{k=1}^\infty \epsilon^k \, \Delta^{(k)}_{q[\mathfrak{n}]}(L) \bigg ]^m \,,
\label{eq:tau}
\end{equation}
where the above analysis establishes
\begin{equation}
T_{\frak n, - 1}(\vec d, L) = - \frac{ g_{\frak n} E^{(0)}_\gothicn (L)^2 }{ 4 \omega_{\bfnu_\gothicn} \omega_{\vec d-\bfnu_\gothicn} } \,.
\end{equation}
The definitions for $G_{\frak n, m}$, $B_{\frak n, m}$ and $T_{\frak n, m}$ can be read off from matching the definitions of the underlying functions and the expansions. The least trivial of these is $T_{\frak n, m}$ which can be defined as
\begin{equation}
T_{\frak n,m}(\vec d, L) \equiv \bigg ( \frac{\partial}{\partial q^2} \bigg )^{\!\! m} \bigg [ \frac{ g_{\frak n} E^{(0)}_\gothicn (L)^2 }{ 4 \omega_{\bfnu_\gothicn} \omega_{\vec d-\bfnu_\gothicn} } \Big [ q^{2} - q^{(0)}_{\frak n}(L)^2 \Big ]^{-1} - \sum_{\vec v \in \vec S_\mathfrak{n}} \Big [ q^{2} - \Gamma(\vec v \vert q, \vec d, L) \Big ]^{-1} \bigg ] \,.
\end{equation}
This collective set of expansions, in particular eqs.~\eqref{eq:boost}, \eqref{eq:beta} and \eqref{eq:tau}, define the main result of this subsection.
This completes the discussion of the general strategy for deriving the expansion about non-interacting energies, for both zero and non-zero momenta in the finite-volume frame as well as for all excited states, in the $S$-wave only truncation.
However, to make such an expansion well-defined, one requires an ansatz for $p^\star \cot \delta_0(p^\star)$ as well as a power-counting scheme, as we discuss in the following section.

\subsection{Power-counting schemes}
\label{subsec:PowerCS}

In the expansion of the finite-volume energy, eq.~\eqref{eq:EnergyExpansion}, the $\epsilon$ parameter used to organize the expansion is ambiguous. For the rest-frame ground-state energy, $1/L$ serves as a natural parameter and, after working out the details of the expansion, one identifies $a_0/L$ as the relevant dimensionless quantity to identify with $\epsilon$.

However, for excited states and for both the ground and excited states in moving frames, it is not always most useful to expand in powers of $1/L$. This is because the non-interacting energy $E_{\mathfrak{n}}^{(0)}(L)$, defined in eq.~\eqref{eq:E0ndef} above, is $L$-dependent. Of course one can simply expand the difference $E_{\mathfrak{n}}(L) - E_{\mathfrak{n}}^{(0)}(L)$ in powers of $1/L$, but as we have shown in eq.~\eqref{eq:LOGeneralResult}, $E_{\mathfrak{n}}^{(0)}(L)$ also appears as a natural building block at higher orders
and expanding this dependence in powers of $1/L$ significantly reduces the descriptive power of the expansion without really simplifying the result.

Based on these considerations, we have found it most useful to organize the expansion by assigning a power-counting scheme to the parameters entering the scattering amplitude, $\mathcal M_{\ell}$. To this end we first give a generalization of the threshold expansion of eq.~\eqref{eq:pCdThreshold}. For the $\ell^{\text{th}}$ partial wave one can write
\begin{equation}
p^\star \cot \delta_{\ell}(p^\star) = - \frac{1}{a_{\ell}}\left(\frac{1}{ p^\star} \right)^{\!\! 2\ell} \bigg [ 1 - \frac{r_\ell a_\ell}{2} p^{\star2} + O(p^{\star 4}) \bigg ] \,.
\label{eq:EffRangeExp}
\end{equation}
Note that, in this convention, the scattering length and the effective range have dimensions that depend on the partial wave of interest, namely
\beq
[a_\ell] = [E]^{-2\ell-1} \qquad [r_\ell] = [E]^{2\ell-1} \, .
\eeq
It is also possible to expand $p^\star \cot \delta_{\ell}(p^\star)$ around different values of $p^\star$, not just about threshold.

As a first example we define a \emph{threshold scheme} via
\begin{equation}
a_{\ell} = \mathcal O(\epsilon^{2 \ell + 1}) \,, \qquad \qquad a_{\ell}^2 r_{\ell} = \mathcal O(\epsilon^{2 \ell + 3}) \,,
\end{equation}
with higher orders dictated by assigning $\epsilon^{2 \ell + 1 + n}$ to the $(p^\star)^n$ coefficient in the effective range expansion of $\tan \delta_{\ell}(p^\star)/p^{\star}$. This scheme is useful when the contributions from higher-partial waves are suppressed. For example for identical particles with $\boldsymbol P = [000]$ and $\Lambda = A_{1g}$, the lowest-lying partial wave contamination arise from $\ell = 4$ and thus appears in this counting at $\mathcal O(\epsilon^9)$, corresponding to the $1/L^9$ scaling identified in ref.~\cite{Luscher:1990ux}.

An alternative approach is to expand $p^\star \cot \delta_{\ell}(p^\star)$ about the value of the non-interacting energy. This approach, referred to below as the \emph{weakly-interacting scheme} directly corresponds to eq.~\eqref{eq:PhaseExp} above. Extending this to all partial waves, we write
\begin{equation}
\mathcal A^{\ell}_m = \mathcal O(\epsilon^{m}) \,,
\end{equation}
where
\begin{equation}
\mathcal A^{\ell}_m \equiv \left(\frac{\partial}{\partial p^{\star 2}}\right)^{\!\!m} \frac{ \tan \delta_\ell(p^\star)}{p^\star} \bigg \vert_{p^{\star2} = (2 \pi/L)^2 q_\mathfrak{n}^{(0)}(L)^2} \,.
\label{eq:KFuncDef}
\end{equation}
In contrast to the threshold scheme, the weakly-interacting scheme does not assume suppression of higher partial waves. This power-counting is appropriate, for example, if we are expanding a highly excited state for which $ (2 \pi/L)^2 q_\mathfrak{n}^{(0)}(L)^2$ is order one. As we discuss in the following section, in this regime all partial waves contribute at leading order. Independent of counting $\mathcal A^{\ell}_0$ as leading-order, one must specify the value of this quantity to give numerical results. Here it may be useful to re-expand $\tan \delta_\ell(p^\star)$ about threshold, to identify
\begin{equation}
\mathcal A^{\ell}_0 = - a_{\ell} \big (p_{\gothicn}^{(0)\star} \big )^{2 \ell} + \mathcal O\big [ (p_{\gothicn}^{(0)\star} )^{2 \ell+2} \big ]\,,
\end{equation}
where $ (p_{\frak n}^{(0)\star} )^{2} = (2 \pi/L)^2 q_{\frak n}^{(0)}(L)^2$. This is the notation used in the following section, but one can easily make the substitution $ - a_{\ell} (p_{\gothicn}^{(0)\star} )^{2 \ell} \to \mathcal A^{\ell}_0 $ to describe a more general phase shift.

\begin{table}[t]
\begin{center}
\begin{tabular}{c | c }
\ \ $n$ \ \ & \ \ $B_{\gothicn, 0}$ \\ \hline \hline
0 & $\phantom{+}8.9136$ \\
1 & $\phantom{+}1.2100$ \\
2 & $\phantom{+}5.0961$ \\
3 & $\phantom{+}6.7745$ \\
4 & $-9.5381$ \\
5& $-7.0197$ \\
6 & $\phantom{+}23.201$
\end{tabular}
\caption{Evaluation of the coefficient $B_{\gothicn, 0}$ for the first few excited states with $\vec P = [000]$.\label{tab:Bn0}}
\end{center}
\end{table}

Working to higher orders adds significant complication in general, both due to the fact that the algebra becomes more involved and also because infinite sums appear in the energy expressions beyond leading order. For example, for $\boldsymbol P =[000]$, the next-to-next-to-leading-order (NNLO) energy for any excited state in the threshold scheme, is
\beq
E_\gothicn(L)
&\equiv& E^{(0)}_\gothicn (L) + \epsilon \,g_\gothicn\frac{8 \pi a_0}{E_{\mathfrak n}^{(0)}(L) L^3} + \epsilon^2\, g_\gothicn \frac{8 a_0^2} {E^{(0)}_\gothicn(L) L^4} \left(B_{\gothicn, 0}-\frac{4\pi^2 g_\gothicn}{E^{(0)}_\gothicn(L)^2 L^2} \right)+\CO(\epsilon^3) \,,
\label{eq:NNLOCoM}
\eeq
where
\beq
\label{eq:Bdef}
B_{\gothicn, 0} & \equiv - \displaystyle \lim_{s \to -1} \sum_{\vec v \notin \vec S_\mathfrak{n}} \Big [ q^{(0)2}_{\frak n} - \Gamma(\vec v \vert q^{(0)}_{\frak n}, \vec d, L) \Big ]^{\! s} \,.
\eeq
As with the leading-order expression of \eq~\eqref{eq:LOGeneralResult}, this result only holds for states that are not accidentally degenerate.
Numerical values of $B_{\gothicn, 0}$ for $\boldsymbol P=[000]$ are given in table~\ref{tab:Bn0}. In the case of non-zero momentum in the finite-volume frame, this quantity inherits an $L$ dependence.
Extending the calculation to moving frames or higher orders in the CoM frame becomes quite cumbersome, but can be readily automated using any computer algebra system.

\subsection{Higher partial waves: Without accidental degeneracies}
\label{subsec:HPW}

To go beyond the $S$-wave-only truncation of sections~\ref{sec:sWaveOnly} and \ref{sec:sWaveOnlyAllOrders}, we return to the full quantization condition, summarized by eq.~\eqref{eq:QC}, and evaluate the determinant for a given set of quantum numbers and for some nonzero value of $\ell_{\sf max}$. For example, setting $\vec P = [000]$, $\Lambda = A_{1g}$ and $\ell_{\sf max} = 4$, one reaches a determinant of a two-dimensional matrix that can be written as
\begin{equation}
\det \bigg [ \begin{pmatrix}
p \cot \delta_0(p) & 0 \\
0 & p \cot \delta_4(p)
\end{pmatrix}
-
\left(\begin{array}{cc}
f_{00}(q, L) \ & \ f_{40}(q, L) \\
f_{40}(q, L) \ & \ f_{44}(q, L)
\end{array}\right)
\bigg ]= 0
\label{eq:TwoByTwoMatrix} \,,
\end{equation}
where $f_{\ell \ell'}$ is given by projecting $F_{\ell m; \ell' m'}$ into the trivial irrep of the octahedral group, taking the real part and rescaling by the factor used to relate $F_{00,00}$ to $f(q, \boldsymbol d, L)$ in eq.~\eqref{eq:fdef}. The trivial component here matches that of the $S$-wave only relation [$f_{00}(q,L)=f(q,\vec 0 , L)$] while the others are described in appendix \ref{app:fvsg}. The reduction to a two-dimensional matrix arises from the fact that $f_{\ell \ell'} = 0$ whenever $\ell$ or $\ell'$ is equal to 1, 2 or 3. Thus, these entries can be dropped without affecting the resulting determinant.

Exactly as for $f_{00}$, the additional $f_{\ell \ell'}$ functions can be expanded about $q^2 = q_{\frak n}^{(0)}(L)^2$ and the condition of vanishing determinant can be solved order by order for a given power-counting scheme. For the threshold scheme, one has to work to very high order to first see the effect of $\ell = 4$. By contrast, in the weakly-interacting scheme, the leading shift for the first excited state (with $q_{\frak n}^{(0) }(L)^2 = 1$) already depends on both partial waves.\footnote{The effect of higher partial waves on the ground state is suppressed in both schemes. This is because the shift away from threshold, required to induce dependence on $\delta_{\ell}$ with $\ell > 0$, arises only due to the interactions. Thus the leading higher-partial waves appear in a product with parameters describing the $S$-wave interaction.} Evaluating eq.~\eqref{eq:TwoByTwoMatrix} and expanding, one reaches
\beq
\Delta^{(1)}_{E[\mathfrak{n}]}=\frac{48\pi }{E^{(0)}_{\frak n}(L) L^3}\left(a_0+\frac{21}{4} a_4 \big (p_{\gothicn}^{(0)} \big )^8\right) \,,
\label{eq:CoML4}
\eeq
where $p_{\gothicn}^{(0)} \equiv 2\pi q_\gothicn^{(0)}/L = 2\pi/L$.

In fact, in the weakly-interacting power-counting scheme, all higher partial waves should be included at leading order. This converts \eq~\eqref{eq:TwoByTwoMatrix} to an infinite-dimensional matrix, which can be studied order by order in the expansion. To see how this works at leading order, note that $f_{ \ell \ell'}$ is given by
\begin{align}
\nonumber
f_{\ell \ell'}(q, L) & \equiv - 8 \pi E \, \mathbb P_{A_{1g},\ell m} \, \mathbb P_{A_{1g},\ell' m'} \, \text{Re}
\lim_{\alpha \to 0^+} \bigg [\frac{1}{L^3} \sum_{\vec k} - \int_{\vec k} \bigg ] \frac{\mathcal Y_{\ell m}( {\vec k}) \mathcal Y^*_{\ell' m'}( {\vec k}) e^{- \alpha ( k^{ 2} - p^{ 2})} }{(2 \omega_{\vec k})^2 (E - 2 \omega_{\vec k} + i \epsilon)} \,, \\
& = - \frac{8 \pi g_{\frak n}}{E^{(0)}_{\frak n}(L) L^3} \frac{M^{\frak n}_{\ell \ell'} }{ \epsilon \Delta^{(1)}_{E[\mathfrak{n}]}} + \mathcal O(\epsilon^0) \,,
\label{eq:littlefLOallPW}
\end{align}
where the definition is taken from appendix~\ref{app:FFunction} and $\mathcal Y_{\ell m}( {\vec k} ) = \sqrt{4 \pi} (k/p)^{\ell} Y_{\ell m}(\hat {\vec k})$. We introduce $\mathbb P_{A_{1g},\ell m}$, which projects the $\ell^{\text{th}}$ partial wave into the trivial irrep and
\begin{align}
\label{eq:Mdef}
M^{\frak n}_{\ell \ell'} & = 4 \pi \mathbb P_{A_{1g},\ell m} \, \mathbb P_{A_{1g},\ell' m'} \frac{1}{\vert O_h \vert} \sum_{ R \in O_h } Y_{\ell m}(R \cdot \hat {\vec \nu}_{\frak n}) Y^*_{\ell' m'}(R \cdot \hat {\vec \nu}_{\frak n}) \,.
\end{align}
As we show in appendix \ref{app:fvsg}, this can be rewritten as
\begin{equation}
\label{eq:MrankOne}
M^{\frak n}_{\ell \ell'} = \sqrt{ \mathcal P^{\frak n}_\ell } \, \sqrt{\mathcal P^{\frak n}_{\ell '} }\,,
\end{equation}
where
\begin{equation}
\label{eq:Pellresult}
\mathcal P^{\frak n}_{\ell} = (2 \ell+1) \frac{1}{\vert O_h \vert} \sum_{R \in O_h} P_{\ell}(\hat {\vec \nu}_{\frak n} \cdot R \cdot \hat {\vec \nu}_{\frak n}) \,,
\end{equation}
and $ P_{\ell}(\hat {\vec \nu}_{\frak n} \cdot R \cdot \hat {\vec \nu}_{\frak n}) = P_{\ell}(\cos \theta)$ is the $\ell^{\text{th}}$ Legendre polynomial. The first few non-zero values of $\mathcal P^{\frak n}_{\ell}$ are summarized in table \ref{tab:Pell}. The key point is that $M^{\frak n}_{\ell \ell'}$ is a rank-one matrix, which allows one to analytically evaluate the leading-order determinant. Note that this is only valid when one is expanding about a state that does not exhibit accidental degeneracy. The latter case is discussed in detail in the following subsection.

\renewcommand{\arraystretch}{1.3}
\begin{table}[t]
\begin{center}
$\boldsymbol P = [000] $

\begin{tabular}{c | c c | c c | c c}
\ \ $\ell$ \ \ &\multicolumn{2}{c|}{ \ \ $q^{(0)}_{\frak n}(L)^2 = 1$ \ \ }&\multicolumn{2}{c|}{ \ \ $q^{(0)}_{\frak n}(L)^2 = 2$ \ \ }&\multicolumn{2}{c}{ \ \ $q^{(0)}_{\frak n}(L)^2 = 3$ \ \ } \\ \hline \hline
0 & \multicolumn{2}{c|}{ $1$} & \multicolumn{2}{c|}{ $1$} & \multicolumn{2}{c}{ $1$} \\
4 &$ \frac{21}{4}$ & 5.25000 &$ \frac{21}{64}$ & 0.32813 &$ \frac{7}{3}$ & 2.33333 \\
6 &$ \frac{13}{8}$ & 1.62500 &$ \frac{2197}{512}$ & 4.29102 &$ \frac{416}{81}$ & 5.13580 \\
8 &$ \frac{561}{64}$ & 8.76563 &$ \frac{45441}{16384}$ & 2.77350 &$ \frac{187}{243}$ & 0.76955 \\
10 &$ \frac{455}{128}$ & 3.55469 &$ \frac{455}{131072}$ & 0.00347 &$ \frac{58240}{6561}$ & 8.87670 \\
12 &$ \frac{18575}{1536}$ & 12.0931 &$ \frac{56638775}{6291456}$ & 9.00249 & $ \frac{763975}{177147}$ & 4.31266 \\
14 &$ \frac{17255}{3072}$ & 5.61686 &$ \frac{136676855}{50331648}$ & 2.71553 & $ \frac{1104320}{177147}$ & 6.23392 \\
16 &$ \frac{251009}{16384}$ & 15.3204 &$ \frac{3525395489}{1073741824}$ & 3.28328 & $ \frac{6464161}{531441}$ & 12.1635 \\
18 &$ \frac{254227}{32768}$ & 7.75839 &$ \frac{83891347603}{8589934592}$ & 9.76624 & $ \frac{170017664}{43046721}$ & 3.94961 \\
20 &$ \frac{2422567}{131072}$ & 18.4827 &$ \frac{922978270207}{137438953472}$ & 6.71555 & $ \frac{1881436823}{129140163}$ & 14.5690 \\
\end{tabular}
\caption{Numerical values for the $\mathcal P^{\frak n}_{\ell}$ coefficients, giving the contribution of the $\ell^{\text{th}}$ partial wave to the first few excited states with $\Lambda = A_{1g}$ and $\boldsymbol P = [000]$.\label{tab:Pell}}
\end{center}
\end{table}
\renewcommand{\arraystretch}{1}

Combining eqs.~\eqref{eq:QC}, \eqref{eq:TwoByTwoMatrix}, \eqref{eq:littlefLOallPW} and \eqref{eq:MrankOne}, we reach a determinant of the form
\begin{equation}
\det \! \bigg [ \underline{p \cot \delta(p)} + \frac{8 \pi g_{\frak n}}{E^{(0)}_{\frak n}(L) L^3} \frac{1}{\epsilon \Delta^{(1)}_{E[\frak n]}} \sqrt{\mathcal P^{\frak n}} \otimes \sqrt{ \mathcal P^{\frak n}} \bigg ] = 0 \, \,,
\end{equation}
where $ \underline{p \cot \delta(p)}$ is a diagonal matrix populated by $p \cot \delta_{\ell}(p)$. This can be rewritten as the eigenvalue equation
\begin{equation}
\bigg [ 1 + \frac{8 \pi g_{\frak n}}{E^{(0)}_{\frak n}(L) L^3} \frac{1}{\epsilon \Delta^{(1)}_{E[\frak n]}} \sqrt{\frac{\tan \delta(p)}{p}} \sqrt{\mathcal P^{\frak n}} \otimes \sqrt{ \mathcal P^{\frak n}} \sqrt{\frac{\tan \delta(p)}{p}} \bigg ] \boldsymbol {\mathcal E} = 0 \, \,,
\end{equation}
at which point one can read off $ \boldsymbol {\mathcal E} = \sqrt{\frac{\tan \delta(p)}{p}} \sqrt{ \mathcal P^{\frak n}} $ and thus
\begin{equation}
1 + \frac{8 \pi g_{\frak n}}{E^{(0)}_{\frak n}(L) L^3} \frac{1}{\epsilon \Delta^{(1)}_{E[\frak n]}} \bigg ( \sqrt{ \mathcal P^{\frak n}} \cdot \frac{\tan \delta(p)}{p} \cdot \sqrt{\mathcal P^{\frak n}} \bigg ) = 0 \, \,.
\end{equation}

Substituting the leading-order relation $p \cot \delta_{\ell}(p) = - \big [ a_\ell (p_{\gothicn}^{(0)} )^{2 \ell} \big ]^{-1}$ and solving for the energy shift, one finally reaches
\begin{align}
\label{eq:Delta1AllEll}
\Delta^{(1)}_{E[\mathfrak{n}]}&=\frac{8 \pi g_{\frak n} }{E^{(0)}_{\frak n}(L) L^3} \sum_{\ell = 0}^\infty \mathcal P^{\frak n}_{\ell} \, a_\ell \, \big (p_{\gothicn}^{(0)} \big )^{2 \ell} \,.
\end{align}
A similar result to this was already derived in ref.~\cite{Luscher:1986pf}, with $E_{\frak n}^{(0)}(L)$ expanded about $L = \infty$. As mentioned above, this result can be re-expressed using $ - a_{\ell} (p_{\gothicn}^{(0)} )^{2 \ell} \to \mathcal A^{\ell}_0 $, in case one has a better description of $\tan \delta_{\ell} (p_{\gothicn}^{(0)})$ than is given by its leading-order threshold expansion.

This can be readily generalized to non-zero $\boldsymbol P$, taking advantage of the fact that $f_{\ell \ell'}(q, \boldsymbol d, L)$ is also a rank-one matrix at leading order, when expanded about a non-degenerate state in the moving frame. One finds
\begin{align}
\nonumber
f_{\ell \ell'}(q, \boldsymbol d, L) = - \frac{E^{(0)}_\gothicn (L) }{ 4 \omega_{\bfnu_\gothicn} \omega_{\vec d-\bfnu_\gothicn} }\frac{8 \pi g_{\mathfrak{n}}}{ \gamma_\mathfrak{n}^{(0)} L^3} \frac{ \sqrt{ \mathcal P^{\frak n}_{\ell} } \sqrt{ \mathcal P^{\frak n}_{\ell'} } }{ \epsilon \Delta^{(1)}_{E[\mathfrak{n}]}} + \mathcal O(\epsilon^0) \,,
\end{align}
from which follows
\begin{align}
\label{eq:Delta1AllEllMF}
\Delta^{(1)}_{E[\mathfrak{n}]}&= \frac{E^{(0)}_\gothicn (L) }{ 4 \omega_{\bfnu_\gothicn} \omega_{\vec d-\bfnu_\gothicn} }\frac{8 \pi g_{\mathfrak{n}}}{ \gamma_\mathfrak{n}^{(0)} L^3} \sum_{\ell = 0}^\infty \mathcal P^{\frak n}_{\ell} \, a_\ell \, \big (p_{\gothicn}^{(0)\star} \big )^{2 \ell} \,,
\end{align}
where
\begin{equation}
\label{eq:PellresultMF}
\mathcal P^{\frak n}_{\ell} = (2 \ell+1) \frac{1}{\vert \LGP \vert} \sum_{R \in \LGP} P_{\ell}(\hat {\vec \nu}_{\frak n}^\star \cdot R \cdot \hat {\vec \nu}_{\frak n}^\star) \,.
\end{equation}
Here $\LGP$ is the relevant little group for the indicated total momentum, i.e.~the subgroup of $O_h$ under which $\vec P$ is left invariant. In general, $\mathcal P^{\frak n}_{\ell}$ depends on $\gamma^{(0)}_\gothicn$, the Lorentz boost factor. For example, for $\boldsymbol P = [001]$ and $\bfnu_\gothicn = (0, 1,1)$, the lowest lying components are
\begin{equation}
\mathcal P^{\frak n}_{2} = \frac{5 (1-2 \gamma ^2 )^2}{ (1+4 \gamma ^2 )^2} \,, \qquad \mathcal P^{\frak n}_{4} = \frac{9 (1- 24 \gamma^2 + 156 \gamma^4 - 144 \gamma^6+ 176 \gamma^8 )}{ (1+4 \gamma ^2 )^4}\,,
\end{equation}
where we have abbreviated $\gamma = \gamma^{(0)}_\gothicn$. The gamma dependence translates to a dependence on $mL$ at fixed $\frak n$ and $\boldsymbol d$.

While the numerical values of $\mathcal P^{\frak n}_{\ell}$ for a given $mL$ can be easily determined, the expressions in terms of $\gamma^{(0)}_{\frak n}$ become very complicated. Therefore, instead of giving additional analytic expressions, we provide numerical values for several states and total momenta at $mL = 4$ and $6$. These are collected in tables~\ref{tab:Pell001}, \ref{tab:Pell011}, \ref{tab:Pell111} and \ref{tab:Pell002}, which give results for $\boldsymbol P = [001]$, $[011]$, $[111]$ and $[002]$ respectively.
Cases do arise for which $\bfnu^\star_\gothicn$ and $\mathcal P^{\frak n}_{\ell}$ do not depend on $\gamma_{\frak n}^{(0)}$. These include trivial examples, in which $\bfnu_\gothicn$ is parallel to $\bfd$, as well as more interesting cases in which a cancellation occurs in the definition of the boost. See tables \ref{tab:Pell011} and \ref{tab:Pell002} for detailed examples.

\renewcommand{\arraystretch}{1.2}
\begin{table}[t]
\begin{center}
$\boldsymbol P = [001] $
\begin{tabular}{c | c c | c c | c c | c c }
\ \ & \multicolumn{2}{c}{$\bfnu_\gothicn = {(0,0,0)}$} & \multicolumn{2}{c}{$\bfnu_\gothicn = {(0,1,1)}$}& \multicolumn{2}{c}{$\bfnu_\gothicn = {(1,1,1)}$}& \multicolumn{2}{c}{$\bfnu_\gothicn = {(0,1,2)}$} \\\hline \hline
\ \ $m L$&$ 4$& $ 6$ &$ 4$& $ 6$ &$ 4$& $ 6$ &$ 4$& $6$ \\ \hline
\diagbox{$\ell$}{$\!\!\!\!\gamma^{(0)}_\gothicn$}& $1.19626$& $1.10634$& $1.07431 $& $1.05691 $& $1.04639 $ & $1.03900$ & 1.03504 & 1.03055 \\ \hline \hline
0 & \multicolumn{2}{c|}{1} & \multicolumn{2}{c|}{1} & \multicolumn{2}{c|}{1} & \multicolumn{2}{c}{1} \\
2 & \multicolumn{2}{c|}{5}& 0.27129 & 0.25468 & 0.59963 & 0.59284 & 1.33212 & 1.34687 \\
4 & \multicolumn{2}{c|}{9} & 2.45998 & 2.43769 & 3.20612 & 3.18518 & 0.27682 & 0.26394 \\
6 & \multicolumn{2}{c|}{13}& 2.61771 & 2.77009 & 0.40378 & 0.42879 & 3.66277 & 3.64266 \\
8 & \multicolumn{2}{c|}{17} & 5.61785 & 5.46126 & 5.77910 & 5.81357 & 5.75364 & 5.75043 \\
10 & \multicolumn{2}{c|}{21} & 5.34109 & 5.30821 & 5.63594 & 5.67937 & 6.21060 & 6.25678 \\
12 & \multicolumn{2}{c|}{25} & 3.22650 & 3.27018 & 7.92711 & 7.79359 & 2.73187 & 2.79102 \\
14 & \multicolumn{2}{c|}{29} & 11.1664 & 11.4538 & 5.60555 & 5.58544 & 7.11547 & 6.90712 \\
16 & \multicolumn{2}{c|}{33}& 6.01485 & 5.62644 & 6.05785 & 6.07222 & 13.4850 & 13.5607 \\
18 & \multicolumn{2}{c|}{37} & 10.8386 & 10.6632 & 10.9403 & 11.1847 & 6.22219 & 6.46048 \\
20 & \multicolumn{2}{c|}{41} & 6.67668 & 7.40405 & 10.7249 & 10.6259 & 7.20059 & 6.96068
\end{tabular}
\caption{Numerical values of the $\mathcal P^\gothicn_{\ell}$ coefficients, giving the leading-order contribution of the $\ell^{\text{th}}$ partial wave to the first few low-lying states for $\Lambda = A_{1}$ and $\boldsymbol P = [001]$. The coefficients generally depend on the Lorentz boost factor $\gamma_\gothicn^{(0)}$ but with exceptions, such as $\vec \nu_{\frak n} = (0,0,0)$ here. For this special case one simply has $\mathcal P^\gothicn_{\ell} = (2 \ell +1)$. In the general case $\mathcal P^\gothicn_{\ell}$ depends on $mL$ and here we evaluate the coefficients for $mL =4$ and $mL = 6$ as indicated.\label{tab:Pell001}}
\end{center}
\end{table}
\renewcommand{\arraystretch}{1}

\renewcommand{\arraystretch}{1.2}
\begin{table}[t]
\begin{center}
$\boldsymbol P = [011] $
\begin{tabular}{c | c c | c c | c c | c c }
\ \ & \multicolumn{2}{c}{$\bfnu_\gothicn = {(0,0,0)}$} & \multicolumn{2}{c}{$\bfnu_\gothicn = {(1,1,1)}$}& \multicolumn{2}{c}{$\bfnu_\gothicn = {(1,1,0)}$}& \multicolumn{2}{c}{$\bfnu_\gothicn = {(0,1,2)}$} \\\hline \hline
\ \ $m L$&$ 4$& $ 6$ &$ 4$& $ 6$ &$ 4$& $ 6$ &$ 4$& $6$ \\ \hline
\diagbox{$\ell$}{$\!\!\!\!\gamma^{(0)}_\gothicn$}& $1.31075$& $1.18046$& $1.13063 $& $1.10236 $& $1.12358 $ & $1.09877$ & 1.09259 & 1.07675 \\ \hline \hline
0 & \multicolumn{2}{c|}{1} & \multicolumn{2}{c|}{1} & \multicolumn{2}{c|}{1} & \multicolumn{2}{c}{1} \\
2 & \multicolumn{2}{c|}{5} & 1.96835 & 1.90202 & \multicolumn{2}{c|}{$ \frac{5}{3}\sim 1.66667$} & 2.34477 & 2.38673 \\
4 & \multicolumn{2}{c|}{9} & 3.67737 & 3.84859 & \multicolumn{2}{c|}{$ \frac{41}{9} \sim 4.55556$ } & 2.93896 & 2.88123 \\
6 & \multicolumn{2}{c|}{13} & 8.60380 & 8.51008 & \multicolumn{2}{c|}{$ \frac{1885}{243}\sim 7.75720$ }& 8.35018 & 8.25431 \\
8 & \multicolumn{2}{c|}{17} & 6.65130 & 6.38607 & \multicolumn{2}{c|}{$ \frac{4505}{729} \sim 6.17970$} & 8.87560 & 9.13679 \\
10 & \multicolumn{2}{c|}{21}& 10.3638 & 11.0137 & \multicolumn{2}{c|}{$\frac{254261}{19683} \sim 12.9178 $} & 7.79505 & 7.70573 \\
12 & \multicolumn{2}{c|}{25} & 14.9275 & 14.3335 & \multicolumn{2}{c|}{$ \frac{5956625}{531441}\sim 11.2084 $} & 14.5092 & 14.1373 \\
14 & \multicolumn{2}{c|}{29} & 11.4305 & 11.3172 & \multicolumn{2}{c|}{$ \frac{7345903}{531441}\sim 13.8226 $} & 15.8750 & 16.3929 \\
16 & \multicolumn{2}{c|}{33} & 17.7535 & 18.8254 & \multicolumn{2}{c|}{$ \frac{30465875}{1594323}\sim 19.1090$} & 12.9680 & 13.0249 \\
18 & \multicolumn{2}{c|}{37} & 20.3293 & 18.8820 & \multicolumn{2}{c|}{$ \frac{1932764005}{129140163}\sim 14.9664 $} & 20.0635 & 19.2635 \\
20 & \multicolumn{2}{c|}{41} & 16.7256 & 17.3828 & \multicolumn{2}{c|}{$ \frac{9057620329}{387420489}\sim 23.3793$} & 23.0663 & 23.7404
\end{tabular}
\caption{As in table \ref{tab:Pell001} but for $\boldsymbol P = [011]$. Note that here two cases arise for which $\mathcal P^\gothicn_{\ell}$ is independent of $mL$, the trivial case with $\bfnu_\gothicn = {(0,0,0)}$ and a more interesting case with $\bfnu_\gothicn = {(1,1,0)}$. \label{tab:Pell011}}
\end{center}
\end{table}
\renewcommand{\arraystretch}{1}

\renewcommand{\arraystretch}{1.2}
\begin{table}[t]
\begin{center}
$\boldsymbol P = [111] $
\begin{tabular}{c | c c | c c | c c | c c }
\ \ & \multicolumn{2}{c}{$\bfnu_\gothicn = {(0,0,0)}$} & \multicolumn{2}{c}{$\bfnu_\gothicn = {(0,0,1)}$}& \multicolumn{2}{c}{$\bfnu_\gothicn = {(0,0,-1)}$}& \multicolumn{2}{c}{$\bfnu_\gothicn = {(0,1,2)}$} \\\hline \hline
\ \ $m L$&$ 4$& $ 6$ &$ 4$& $ 6$ &$ 4$& $ 6$ &$ 4$& $6$ \\ \hline
\diagbox{$\ell$}{$\!\!\!\!\gamma^{(0)}_\gothicn$}& $1.39618$ & $1.23919$ & $1.29171 $& $1.20769 $& $1.13026 $ & $1.10865$ & 1.11783 & 1.10111 \\ \hline \hline
0 & \multicolumn{2}{c|}{1} & \multicolumn{2}{c|}{1} & \multicolumn{2}{c|}{1} & \multicolumn{2}{c}{1} \\
2 & \multicolumn{2}{c|}{5 }& 0.78193 & 0.72807 & 1.59464 & 1.66216 & 0.11819 & 0.10608 \\
4 & \multicolumn{2}{c|}{9 }& 2.37330 & 2.53029 & 0.74414 & 0.67119 & 0.59680 & 0.64024 \\
6 & \multicolumn{2}{c|}{13} & 6.82511 & 6.67527 & 6.46793 & 6.27289 & 2.63788 & 2.57512 \\
8 & \multicolumn{2}{c|}{17} & 2.59823 & 2.50325 & 6.98998 & 7.37341 & 4.73931 & 4.78323 \\
10 &\multicolumn{2}{c|}{ 21} & 8.52575 & 9.01303 & 3.20973 & 3.24944 & 1.19544 & 1.13906 \\
12 &\multicolumn{2}{c|}{ 25} & 9.76633 & 9.05537 & 9.73565 & 9.01811 & 3.28963 & 3.49965 \\
14 & \multicolumn{2}{c|}{29} & 5.74940 & 6.19649 & 13.0258 & 13.5804 & 8.75731 & 8.34118 \\
16 & \multicolumn{2}{c|}{33} & 15.1005 & 15.4210 & 6.78710 & 7.40763 & 2.39753 & 2.75679 \\
18 &\multicolumn{2}{c|}{ 37} & 10.7882 & 9.61485 & 11.7747 & 10.4662 & 6.48135 & 6.55101 \\
20 &\multicolumn{2}{c|}{ 41} & 11.3319 & 12.7961 & 18.7898 & 18.9909 & 6.61792 & 6.18188 \\ \end{tabular}
\caption{As in table \ref{tab:Pell001} but for $\boldsymbol P = [111]$.\label{tab:Pell111}}
\end{center}
\end{table}
\renewcommand{\arraystretch}{1}

\renewcommand{\arraystretch}{1.2}
\begin{table}[t]
\begin{center}
$\boldsymbol P = [002] $
\begin{tabular}{c | c c | c c | c c | c c }
\ \ & \multicolumn{2}{c}{$\bfnu_\gothicn = {(0,0,0)}$} & \multicolumn{2}{c}{$\bfnu_\gothicn = {(0,1,1)}$}& \multicolumn{2}{c}{$\bfnu_\gothicn = {(1,1,2)}$}& \multicolumn{2}{c}{$\bfnu_\gothicn = {(0,1,2)}$} \\\hline \hline
\ \ $m L$&$ 4$& $ 6$ &$ 4$& $ 6$ &$ 4$& $ 6$ &$ 4$& $6$ \\ \hline
\diagbox{$\ell$}{$\!\!\!\!\gamma^{(0)}_\gothicn$}& $1.46576 $& $1.28858 $& $1.30828 $& $1.23412 $& $1.14714 $ & $1.12709$ & 1.21680 & 1.17441 \\ \hline \hline
0 & \multicolumn{2}{c|}{1} & \multicolumn{2}{c|}{1} & \multicolumn{2}{c|}{1} & \multicolumn{2}{c}{1} \\
2 & \multicolumn{2}{c|}{5} &\multicolumn{2}{c|}{$ \frac{5}{4}\sim 1.25$} & 0.03784 & 0.02915 & 0.05480 & 0.08510 \\
4 & \multicolumn{2}{c|}{9} & \multicolumn{2}{c|}{$\frac{99}{16}\sim 6.18750$} & 2.31286 & 2.31576 & 2.25597 & 2.20657 \\
6 & \multicolumn{2}{c|}{13} &\multicolumn{2}{c|}{$ \frac{143}{32} \sim4.46875$} & 4.82646 & 4.90205 & 4.87130 & 4.76372 \\
8 & \multicolumn{2}{c|}{17} & \multicolumn{2}{c|}{$\frac{2771}{256} \sim10.8242$} & 1.70987 & 1.49774 & 1.90006 & 2.41449 \\
10 &\multicolumn{2}{c|}{ 21} &\multicolumn{2}{c|}{$ \frac{4053}{512}\sim7.91602$} & 7.57424 & 7.83763 & 6.77813 & 5.80645 \\
12 &\multicolumn{2}{c|}{ 25} &\multicolumn{2}{c|}{$ \frac{31375}{2048} \sim15.3198$} & 4.88759 & 4.82769 & 6.52095 & 7.59568 \\
14 & \multicolumn{2}{c|}{29} & \multicolumn{2}{c|}{$\frac{46951}{4096} \sim11.4626$} & 7.73567 & 7.34108 & 5.64314 & 5.10550 \\
16 & \multicolumn{2}{c|}{33} &\multicolumn{2}{c|}{$ \frac{1293699}{65536}\sim19.7403$} & 8.16988 & 8.99679 & 9.47708 & 8.99164 \\
18 & \multicolumn{2}{c|}{37} &\multicolumn{2}{c|}{$ \frac{1975097}{131072} \sim15.0688$}& 8.99199 & 8.14196 & 10.1730 & 11.5729 \\
20 & \multicolumn{2}{c|}{41} & \multicolumn{2}{c|}{$\frac{12641653}{524288} \sim24.1120$}& 11.4709 & 11.7618 & 6.75650 & 5.17772 \end{tabular}
\caption{As in table \ref{tab:Pell001} but for $\boldsymbol P = [002]$. As with table \ref{tab:Pell011} a non-trivial case of $mL$ independence arises. \label{tab:Pell002}}
\end{center}
\end{table}
\renewcommand{\arraystretch}{1}

\subsection{Higher partial waves: Including accidental degeneracies}
\label{subsec:AccDeg}

We turn now to the case that the non-interacting states, about which we are expanding, are accidentally degenerate. This occurs whenever one can identify two (or more) values of representative momenta $\boldsymbol \nu_{\frak n,1}$ and $\boldsymbol \nu_{\frak n,2}$ such that the corresponding energies are equal for all $L$,\footnote{Another special case arises when two energies coincide only for a single, finely tuned value of $mL$. Here the non-degenerate expansion can be performed at any $L$ away from the finely tuned point. One then manifestly sees a breakdown in the form of diverging expansion coefficients as $L$ approaches the crossing. An alternative expansion, performed exactly at the degenerate point and treating the accidental degeneracy as described here would resolve the issue and give the correct avoided-level-crossing behavior.} i.e.~
\begin{equation}
E_{\frak n, 1}(L) = E_{\frak n, 2}(L) \,,
\end{equation}
for
\begin{equation}
E_{\frak n, i}(L) \equiv \sqrt{m^2 + (2\pi/L)^2 \boldsymbol \nu^2_{\frak n,i} } + \sqrt{m^2 + (2\pi/L)^2 (\boldsymbol d - \boldsymbol \nu_{\frak n,i})^2 } \,,
\end{equation}
but with the property that $\boldsymbol \nu_{\frak n, 1}$ and $\boldsymbol \nu_{\frak n, 2}$ (and also $\boldsymbol \nu_{\frak n, 1}$ and $\boldsymbol d - \boldsymbol \nu_{\frak n, 2}$) are not related by rotations within $\LGP$. Equivalently, the $\frak{n}$ state is accidentally degenerate whenever the elements of the set $\boldsymbol S_{\frak n}$, defined in eq.~\eqref{eq:E0ndefForn}, are not related by rotations within the little group.

The standard example is the state with $q_\gothicn^{(0)}(L)^2 =9$ and $\vec P = [000]$, which includes $\boldsymbol \nu_{\frak n, 1} = (1,2,2)$ and $\boldsymbol \nu_{\frak n, 2} = (0,0,3)$. In this case, the set $\vec S_\gothicn$ naturally decomposes into two subsets, given by all rotations of the two momentum types, and we say the state is two-fold degenerate. The degeneracy is broken by the lowest non-vanishing angular momentum that couples to the system, $\ell = 4$. Setting $\ell_{\sf max}=4$ and expanding $f_{\ell \ell'}$ to leading order gives
\begin{equation}
\left(\begin{array}{cc}
f_{00}(q, L) \ & \ f_{40}(q, L) \\
f_{40}(q, L) \ & \ f_{44}(q, L)
\end{array}\right)
=
- \frac{240 \pi}{E^{(0)}_{\frak n}(L) L^3} \frac{1 }{ \epsilon \Delta^{(1)}_{E[\mathfrak{n}]}}
\left(\begin{array}{cc} 1 \ & \ \frac{5 }{18} \sqrt{\frac{7}{3}} \\
\frac{5 }{18} \sqrt{\frac{7}{3}} \ & \ \frac{1967}{972}
\end{array}\right) + \mathcal O(\epsilon^0)\,,
\end{equation}
where crucially, and in contrast to the case analyzed in the previous subsection, the leading-order part of $f_{\ell \ell'}$ is no longer rank one and thus multiple solutions arise.
Substituting into eq.~\eqref{eq:TwoByTwoMatrix}, expanding to leading order in $a_0, a_4 = \mathcal O(\epsilon)$ and solving for the leading-order energy shift, one finds
\begin{align}
\label{eq:accdegshift}
\Delta^{(1)}_{E[\mathfrak{n}], \pm}(L)&= \frac{120 \pi}{E_\gothicn^{(0)}(L)L^3} \Big ( \alpha \pm \sqrt{\beta} \, \Big) \,, \\
\alpha & = a_0 +\frac{1967}{972} a_4(p_{\gothicn}^{(0)})^{8} \,, \\
\beta & = a_0^2 - \frac{539}{162} a_0 a_4(p_{\gothicn}^{(0)})^{8} +\left(\frac{1967}{972}\right)^2 a^2_4(p_{\gothicn}^{(0)})^{16} \,.
\end{align}
Expanding the split states about $a_4 = 0$ then yields
\begin{align}
\Delta^{(1)}_{E[\mathfrak{n}], +}(L)&= \frac{240 \pi }{E_\gothicn^{(0)}(L)L^3}a_0 + \mathcal O(a_4) \,, \\
\Delta^{(1)}_{E[\mathfrak{n}], -}(L)&= \frac{35840 \pi }{81 E_\gothicn^{(0)}(L)L^3} a_4 (p_{\gothicn}^{(0)})^{8} + \mathcal O(a_4^2)\,.
\end{align}
It is interesting to note that $\Delta^{(1)}_{E[\mathfrak{n}], +}(L)$ matches the naive non-degenerate result of eq.~\eqref{eq:LOGeneralResult} with $g_{\frak n} = 30$ counting rotations and flips of both $(1,2,2)$ and $(0,0,3)$ within $S_{\frak n}$. (See also table~\ref{tab:SetVectors}.)

A similar calculation can be used to deduce the leading splitting for any accidentally generate state with any value of $\boldsymbol P$. Odd partial waves never contribute, since $\delta_{\ell}(p) = 0$ for identical particles with odd $\ell$. However, in the case of non-zero total momentum, the projected geometric function $f_{\ell \ell'}$ is nonzero for $\ell, \ell' = 2$ and $\delta_{2}(p)$ does enter the energy shift. As with the example presented above, the leading-order energies can always be determined by identifying the value of $\ell_{\sf max}$ required to break the accidental degeneracy and solving the quantization condition truncated to this order.

Rather than give a large number of examples, here we summarize two key observations, proven in appendix \ref{app:ADobservations}:
\begin{itemize}
\item For non-zero $\boldsymbol P$, the accidental degeneracy is sometimes broken only by $\ell=4$, even when $\ell=2$ contributes to the energy shifts. In particular, one can show that for all states with $\boldsymbol P = [00a]$ and $\boldsymbol P = [aaa]$, $\ell_{\sf max}=4$ is required to split degenerate levels. For all other momentum types, the splitting occurs with $\ell_{\sf max}=2$.
\item Determining the leading shift in a power-counting for which a finite set of partial waves are counted as the same order, $a_{\ell} = \mathcal O(\epsilon)$, one generically recovers expressions like eq.~\eqref{eq:accdegshift}, i.p.~not polynomials in the partial-wave coefficients. However, expanding the resulting solutions about $a_{\ell}=0$ for all non-zero $\ell$,
one always recovers one shift of the form
\beq
\Delta^{(1)}_{E[\mathfrak{n}], +}(L) = g_{\gothicn} \frac{E^{(0)}_\gothicn (L) }{ 4 \omega_{\bfnu_\gothicn} \omega_{\vec d-\bfnu_\gothicn} }\frac{8 \pi a_0}{ \gamma_\mathfrak{n}^{(0)} L^3} \,.
\eeq
This is the naive result for non-degenerate states given in eq.~\eqref{eq:accdegshift}.
In the case of a two-fold degeneracy, the other state has a shift of the form
\beq
\Delta^{(1)}_{E[\mathfrak{n}], -}(L) \propto a_{\ell_B} (p_{\gothicn}^{(0)})^{2\ell_B} \,,
\eeq
where $\ell_B$ is the lowest partial wave that is required to split the degeneracy.
\end{itemize}

\section{Numerical checks}
\label{sec:numerical}

\begin{figure}[t]
\centering
\includegraphics[width=\textwidth]{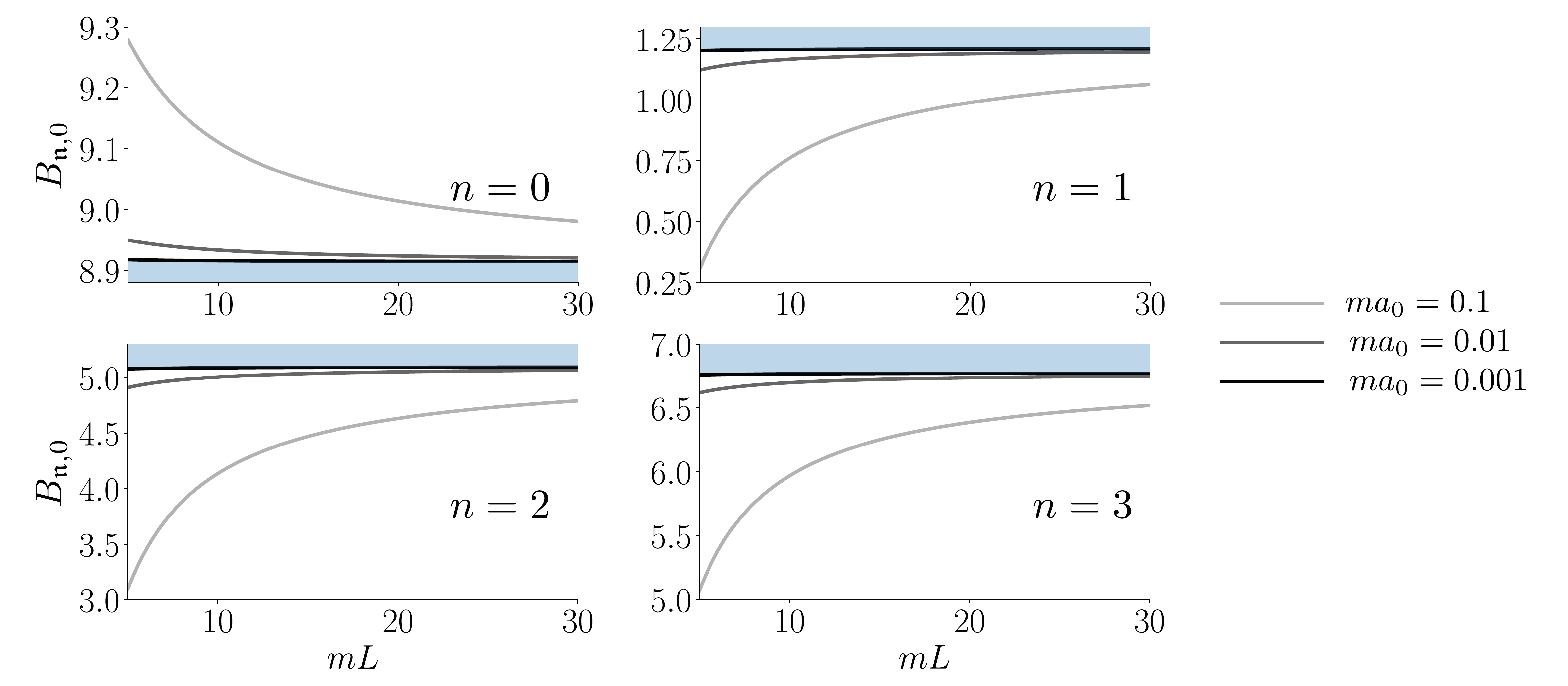}
\caption{Result of equating the numerically determined finite-volume energy, $E_{\frak n}(L)$, to the truncated expansion of eq.~\eqref{eq:NNLOCoM} and solving for the leading coefficient, $B_{\frak n, 0}$. The extraction is performed for the ground state ($n=0$) and the first three excited states ($n=1,2,3$) all for $\vec P = [000]$ and for three different values of $m a_0$, as indicated in the legend. The border between the shaded blue and the unshaded regions indicates the numerical value of $B_{\frak n, 0}$ in each case, given by explicitly evaluating the sum defined in eq.~\eqref{eq:Bdef}. As expected, the extraction from the full $E_{\frak n}(L)$ value approaches the expected value as $mL$ increases and, for fixed $mL$, also as $m a_0$ decreases.}
\label{fig:H0Extraction}
\end{figure}

In the previous section we have laid out a systematic method for expanding a given finite-volume energy in an arbitrary frame to any desired order in a specified power-counting scheme. We have additionally presented the explicit leading-order, and in certain cases higher-order, energy shifts for any non-degenerate state.

The specific results given depend on details of the state and the scheme used. In the threshold scheme, in which the $S$-wave dominates, we give the NLO expression for general momentum $\vec P$ and general excitation $n$ in \eq~\eqref{eq:LOGeneralResult} and the NNLO expression for $\vec P=[000]$ in eq.~\eqref{\SecondMR}. The result in the weakly-interacting scheme is given in \eq~\eqref{\ThirdMR} and contains an infinite sum over angular momentum components with known coefficients. (See also tables \ref{tab:Pell}\,-\,\ref{tab:Pell002}.) Finally, section \ref{subsec:AccDeg} summarizes the leading-order results in the case of an accidentally degenerate state. Here the expansion must be performed, at least initially, treating the degeneracy-breaking partial wave as leading order.

In this section we summarize two tests that verify our methods and provide a cross check on the expressions presented in this work. Both make use of the fact that one can numerically solve the $S$-wave quantization condition for weakly interacting systems and numerically compare to the expanded result.

The first check explicitly addresses \eq~\eqref{eq:NNLOCoM} by taking this result, truncated to the order written, then substituting in the numerical determination of $E_{\frak n}(L)$ and finally solving for an effective $B_{\gothicn, 0}$
\begin{equation}
B_{\frak n, 0}^{\sf eff}(L) \equiv \bigg [ E_\gothicn(L) - E^{(0)}_\gothicn (L) - \,g_\gothicn\frac{8 \pi a_0}{E_{\mathfrak n}^{(0)}(L) L^3} \bigg ] \bigg ( g_\gothicn \frac{8 a_0^2} {E^{(0)}_\gothicn(L) L^4} \bigg)^{\!\!-1} + \frac{4\pi^2 g_\gothicn}{E^{(0)}_\gothicn(L)^2 L^2} \,.
\end{equation}
As we confirm in figure~\ref{fig:H0Extraction}, this quantity asymptotes to the values predicted by eq.~\eqref{tab:Bn0} and listed in table~\ref{tab:Bn0}. The approach to the plateau is consistent with the expected scaling
\begin{equation}
B_{\frak n, 0}^{\sf eff}(L) = B_{\frak n, 0} + \mathcal O(a_0/L) \,.
\end{equation}

\begin{figure}[t]
\centering
\includegraphics[width=\textwidth]{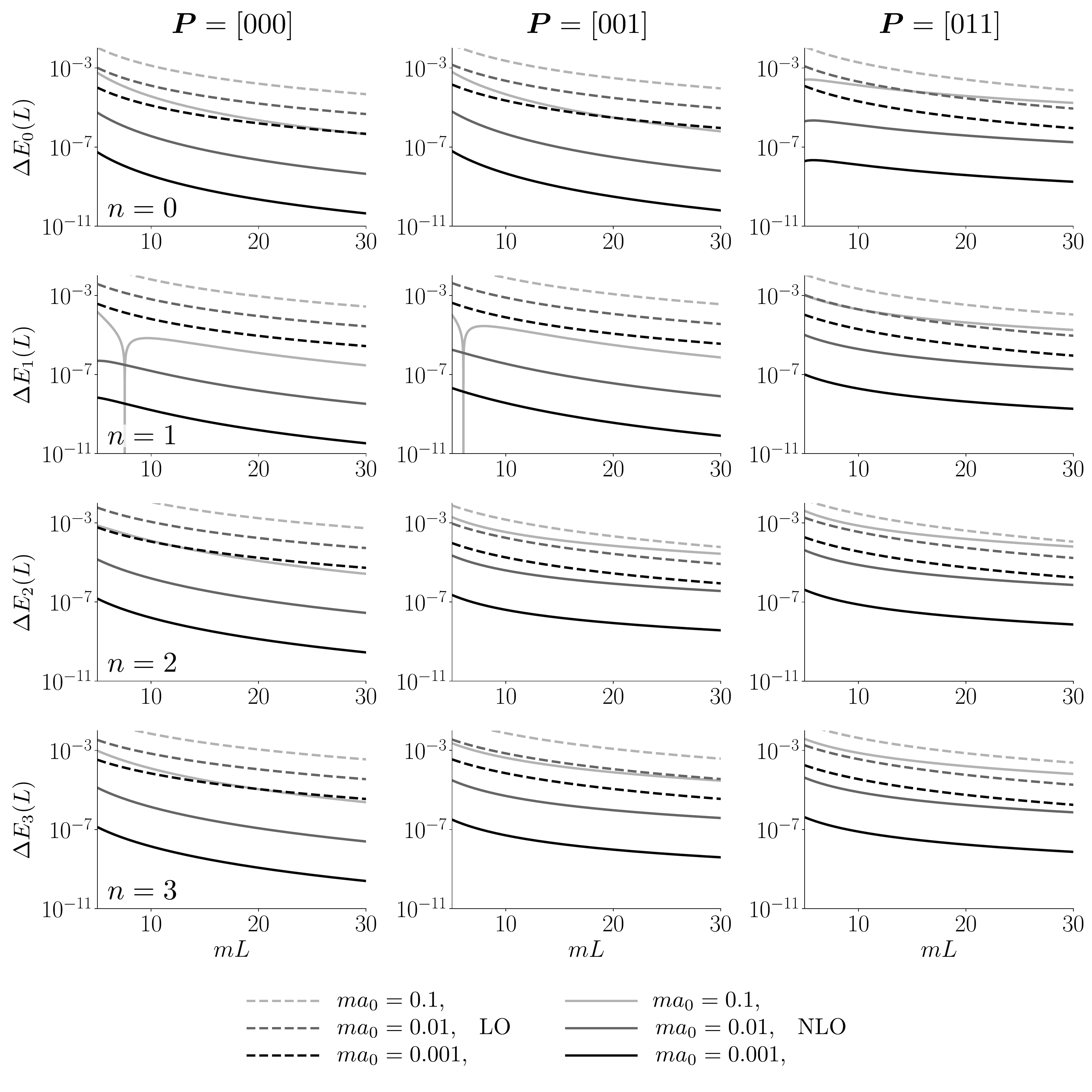}
\caption{The difference between the full finite-volume energy, found by numerically solving the $S$-wave-only quantization condition, and the analytic expressions derived in the manuscript. The three columns show three different momenta, $\vec P = [000], [001],[011]$, and the rows show the ground state (top) and first three excited states (second to fourth rows as labeled). Each panel shows three different scattering length values and two orders of subtraction. As expected, the difference decreases when the subtracted order is increased, and smaller values of $m a_0$ give a smaller residue. }
\label{fig:DeltaE}
\end{figure}

The second check addresses eq.~\eqref{eq:LOGeneralResult} and shows that the difference between our analytic expression and the numerical result from the quantization condition decreases in magnitude as one goes to higher orders in the expansion. This check is performed by subtracting first the LO term (the non-interacting energy) followed by the next-to-leading-order (NLO) correction (the correction proportional to $a_0$). We carry out this analysis for three values of total momenta -- $\boldsymbol P = [000], [001], [011]$ --
and for the four lowest states in each frame.
The results of this comparison are shown in figure~\ref{fig:DeltaE}.

\section{Conclusion and outlook}
\label{sec:Conc}

In this work we have presented analytic expansions of finite-volume two-particle excited states for any value of spatial momentum, $\boldsymbol P$, defined with respect to the finite-volume frame. The results were derived using L{\"u}scher's finite-volume scattering formalism and its extension to the moving frame, and provide intuition for numerical solutions of the latter. In contrast to the rest-frame ground state, for non-zero $\boldsymbol P$ and for excited states, the inverse box length $1/L$ no longer neccesarily defines a useful expansion parameter. This is because the non-interacting energy and the Lorentz boost factor both depend on the volume, and expanding these, as they appear within the energy shift, degrades the range of validity without really simplifying the results. The preferred method is thus to parametrize the scattering amplitude, and to assign a power-counting scheme to the relevant parameters in order to organize the series. This is discussed in section \ref{subsec:PowerCS}.

The general method and concrete results, detailed in section \ref{sec:Derivation}, apply to a single channel of identical spinless particles. Attention is also restricted to finite-volume energies in the trivial irrep of the symmetry group, but we do include the effects of higher orbital angular momenta, which contribute to the trivial irrep due to the reduced rotational symmetry of the cubic box. As is discussed in detail in section \ref{subsec:AccDeg}, non-trivial partial waves play a particularly interesting role in the case of accidentally degenerate states, for which they are required to split the degeneracy.

The main motivation of this work is to establish the method, and the necessary inputs, for an analogous expansion of three-particle finite-volume excited states with generic $\boldsymbol P$. We have derived these results in parallel, by expanding the three-particle quantization condition of refs.~\cite{\LtoK,\KtoM}. The results will be presented in a separate manuscript. The expressions for three-particle states are expected to be particularly useful since the full numerical machinery is significantly more complicated than in the two-particle sector.

In addition to setting the framework for three-particle energies, analytic expressions for two-particle energies are useful in their own right. We have four applications in mind:
First, the results can build intuition on the sensitivity of energy shifts to scattering parameters (e.g. to design a lattice calculation to target a particular scattering observable). This includes basic observations, such as the fact that nonzero total momentum generically reduces the energy shifts and that states with high multiplicity in the non-interacting limit have enhanced shifts. Second, the expansions may be used to understand volume effects in more complicated lattice quantities, by decomposing the latter in a spectral representation (i.e.~inserting a complete set of finite-volume states and inserting expansions for the energies and matrix elements). This could be instructive for the vector-vector correlator, entering the leading-order hadronic vacuum polarization contribution to the muon's magnetic moment, as well as the smeared spectral functions discussed, e.g.~in refs.~\cite{\smearedSF}. Third, the expansions may be useful in designing efficient root finding in numerical solvers of the full quantization condition. Fourth, and finally, the expansions can give information on the convergence of higher partial waves as they enter the finite-volume energies. This is represented by the tables \ref{tab:Pell}-\ref{tab:Pell002}, which summarize the known geometric part of contribution of the $\ell^{\text{th}}$ partial wave's leading contribution to various energies.

This work clearly opens the door to many generalizations, including expansions of the two-body formalism for non-identical and non-degenerate masses, for multiple channels and for particles with intrinsic spin, as well lifting the restriction to the trivial irrep. Another class of extensions would be to adjust the expansion to accommodate poles in $\tan \delta_{\ell}(p)$, that generically arise in systems with a narrow resonance. In all cases it should be stressed that an expansion can never contain more information than a direct numerical solution of the quantization condition, provided the later is evaluated with the same angular-momentum truncation and the same scattering amplitudes. Nevertheless, the analytic understanding provided by this approach is highly instructive and will be a useful tool on the way to increasingly ambitious multi-particle lattice calculations.

\acknowledgments
We thank Fernando Romero-L{\'o}pez and Steve Sharpe for useful discussions. DMG would like to additionally thank Simon Knapen and Michael Wagman for support and useful discussions, and MTH acknowledges Mattia Bruno for useful comments based on presentations of this work in progress. The work of MTH is supported by UK Research and Innovation Future Leader Fellowship MR/T019956/1.

\bibliographystyle{JHEP}
\bibliography{refs.bib}

\appendix

\section{$F$ function}
\label{app:FFunction}

The geometric function $F$, for a single channel of identical spin-zero particles, is defined as \cite{Luscher:1986pf,Luscher:1990ux,Rummukainen:1995vs,\KSS}
\begin{equation}
\label{eq:Fdef}
F_{\ell' m', \ell m}(E, \vec P, L) = \frac 12 \lim_{\alpha \to 0^+} \bigg [\frac{1}{L^3} \sum_{\vec k} - \int_{\vec k} \bigg ] \frac{\mathcal Y_{\ell' m'}( {\vec k}^\star) \mathcal Y^*_{\ell m}( {\vec k}^\star) e^{- \alpha ( k^{\star 2} - p^{\star 2})} }{2 \omega_{\vec k} 2 \omega_{\vec P - \vec k}(E - \omega_{\vec k} - \omega_{\vec P - \vec k} + i \epsilon)} \,.
\end{equation}
The sum-integral difference is specified using
\begin{equation}
\frac{1}{L^3} \sum_{\vec k} = \frac{1}{L^3} \sum_{\substack{\ \vec k = 2 \pi \vec v/L\\\vec v \in \mathbb Z^3}} \,, \qquad \qquad \qquad \qquad \int_{\vec k} = \int \frac{d^3 \vec k}{(2 \pi)^3} \,,
\end{equation}
where the sum runs over integer-vector multiples of $(2 \pi/L)$ and the integral has the usual momentum-space normalization.
The various factors of $\omega$ are defined via
\begin{equation}
\omega_{\vec k} = \sqrt{m^2 + \vec k^2} \,, \qquad \qquad \qquad \qquad \omega_{\vec P - \vec k} = \sqrt{m^2 + (\vec P - \vec k)^2} \,.
\end{equation}
The CoM frame vector $\vec k^\star$
satisfies
\begin{equation}
{\Lambda^\mu}_\nu(- \vec P/E) \begin{pmatrix} \omega_{\vec k} \\ \vec k \end{pmatrix}^{\! \! \nu} = \begin{pmatrix} \omega_{\vec k}^\star \\ \vec k^\star \end{pmatrix}^{\! \! \mu} \,,
\end{equation}
where ${\Lambda^\mu}_\nu(- \vec P/E)$ is the Lorentz boost with velocity given by the argument, i.e.~the boost for which
\begin{equation}
{\Lambda^\mu}_\nu(- \vec P/E) \begin{pmatrix} E \\ \vec P \end{pmatrix}^{\! \! \nu} = \begin{pmatrix} E^\star \\ \vec 0 \end{pmatrix}^{\! \! \mu} \,.
\end{equation}
We denote the magnitude and direction of $\boldsymbol k^\star$ by $k^\star$ and $\hat {\boldsymbol k}^\star$, respectively, i.e.~$\boldsymbol k^\star = k^\star \hat {\boldsymbol k}^\star$.

The numerator of \eq~\eqref{eq:Fdef} includes the generalized spherical harmonics
\begin{equation}
\mathcal Y_{\ell m}( {\vec k}^\star) = \sqrt{4 \pi} \bigg( \frac{k^\star}{p^\star } \bigg)^{\!\ell} Y_{\ell m}(\theta^\star, \phi^\star) \,,
\end{equation}
where $p^\star$ is defined in terms of $E^\star$ by eq.~\eqref{eq: EnergyMomentumRelation} and the angles defined via\break $\hat {\vec k}^\star = (\sin \theta^\star \cos \phi^\star , \sin \theta^\star \sin \phi^\star, \cos \theta^\star)$. The exponential in the numerator is used to regulate the ultraviolet behavior of the sum and integral individually and, as indicated, the regulator independent definition of $F$ is given by sending $\alpha \to 0^+$. Finally, the $i \epsilon$ pole-prescription in $F$ in inherited from the pole-prescription defining the Feynman diagrams appearing in the definition of $\mathcal M(E^\star)$ and is also required to make the integral well-defined.

As was shown in refs.~\cite{\writingFaszeta}, the $F$-functions can be rewritten in terms of generalized zeta functions. For example
\begin{align}
F_{00,00}(E, \vec P, L) & = \frac 12 \lim_{\alpha \to 0^+} \bigg [\frac{1}{L^3} \sum_{\vec k} - \int_{\vec k} \bigg ] \frac{ e^{- \alpha ( k^{\star 2} - p^{\star 2})} }{2 \omega_{\vec k} 2 \omega_{\vec P - \vec k}(E - \omega_{\vec k} - \omega_{\vec P - \vec k} + i \epsilon)} \,, \\
& = \frac{1}{4 E^\star} \lim_{\alpha \to 0^+} \bigg [\frac{1}{L^3} \sum_{\vec k} \frac{\omega_{\boldsymbol k}^\star}{\omega_{\boldsymbol k}} - \int_{\vec k^\star} \bigg ] \frac{ e^{- \alpha ( k^{\star 2} - p^{\star 2})} }{ p^{\star2} - {\vec k}^{\star2} + i \epsilon } \,, \\
& = \frac{1}{16 \pi E^\star} \lim_{s \to -1} \frac{1}{\gamma(q,\boldsymbol d, L) \, \pi L}\sum_{\vec v \in \mathbb Z^3} \Big [ q^{2} - \Gamma(\vec v \vert q, \vec d, L) \Big ]^{\! s} + i \frac{p^\star}{16 \pi E^\star} \,,
\end{align}
where in the second and third lines we have dropped terms scaling as $e^{- m L}$. This result, together with
\begin{equation}
\mathcal M_{00,00}(E^\star)^{-1} = \frac{p^\star \cot \delta_0(p^\star) - i p^\star}{16 \pi E^\star} \,,
\end{equation}
implies eq.~\eqref{eq:SWaveMaster} of the main text.

\section{Level crossing with two-particle energies}
\label{app:NoLevelCrossing}

In this appendix we show that, while non-interacting two-particle levels never intersect as a function of $mL$ for $\boldsymbol P = [000]$, such crossings do occur for non-zero momenta in the finite-volume frame. We also comment on the consequences of this for the definition of the index $n$, within $\frak n = n, \boldsymbol P, \Lambda$.

The non-crossing for $\boldsymbol P=[000]$ follows immediately from the series of inequalities
\beq
\bfnu_{\gothicn_2}^2 > \bfnu_{\gothicn_1}^2
\ \ \Longrightarrow \ \
m^2 + \frac{4\pi^2}{L^2}\bfnu_{\gothicn_2}^2 > m^2 + \frac{4\pi^2}{L^2}\bfnu_{\gothicn_1}^2
\ \ \Longrightarrow \ \
E_{\gothicn_2} &>& E_{\gothicn_1} \,.
\eeq
To see that this breaks for non-zero $\boldsymbol P$ consider this particular example for $\vec P = [003]$:
\beq
E^{(0)}_{2, [003], A_{1g}} &=& \sqrt{m^2+ 2 \frac{4 \pi^2}{L^2} } +\sqrt{m^2+ 5 \frac{4 \pi^2}{L^2} } \,, \qquad \qquad \Big ( \vec \nu_{\frak n} = (0,1,1) \Big ) \,, \\
E^{(0)}_{3, [003], A_{1g}} &=& m +\sqrt{m^2+ 9 \frac{4 \pi^2}{L^2} } \,, \qquad \qquad \Big ( \vec \nu_{\frak n} = (0,0,0) \Big ) \,,
\eeq
where we have explicitly included the ordering index, determined in the large $mL$ limit. The two energies coincide at $mL = 3\pi/\sqrt{2} \sim 6.7$. This represents a second kind of accidental degeneracy. Although perturbative results, such as the that given in \eq~\eqref{eq:NNLOCoM}, hold on either side of the intersection point, the expansion coefficients become arbitrarily large and diverge as the intersection is approached and thus the expansion breaks down.

Finally, for asymptotically large $mL$ one can use the non-relativistic expansion to show
\beq
\left(\bfnu^2_{\gothicn_2}+\left(\vec d -\bfnu_{\gothicn_2}\right)^2\right) > \left(\bfnu^2_{\gothicn_1}+\left(\vec d -\bfnu_{\gothicn_1}\right)^2 \right)
\ \ \Longrightarrow \ \
E_{\gothicn_2} &>& E_{\gothicn_1} \,, \qquad \Big ( L \to \infty \Big ) \,,
\eeq
i.e.~a definitive ordering is restored. This can be used to unambiguously index the energies in studying expansions where such crossings occur.

\section{Finite-volume symmetry and projectors}
\label{app:fvsg}

\begin{table}
\begin{center}
\begin{tabular}{cccc}
$\vec P$& $\LGP$ & $N_\text{elements}$ & $N_\text{irreps}$ \\ \hline
$[000]$ & $O_h$ & 48 & 10 \\
$[00a]$ & $C_{4v}$ & 8 & 5 \\
$[0aa]$ & $C_{2v}$ & 4 & 4 \\
$[0ab]$ & $C_s$ & 2 & 2 \\
$[aaa]$ & $C_{3v}$ & 6 & 3 \\
$[aab]$ & $C_{s}$ & 2 & 2 \\
$[abc]$ & $C_1$ & 1 & 1
\end{tabular}
\end{center}
\caption{Basic properties of finite-volume little groups for a given $\boldsymbol P$, including the name of $\LGP$, the number of group elements and the number of irreps. For more details see refs.~\cite{\fvsg}.}
\label{tab:FiniteVolSym}
\end{table}

In numerical lattice calculations, one generally considers finite-volume energies in a given irrep of the relevant symmetry group. For a cubic, periodic geometry, the relevant group is determined by the total momentum: For $\boldsymbol P = [000]$ the system is invariant under the elements of $O_h$, the 48-element octahedral group (including parity transformations), while for nonzero $\boldsymbol P$ the invariance is reduced to a subgroup of $O_h$, called the point group or little group and denoted by $\LGP$, built from all elements that do not transform the total momentum. In table~\ref{tab:FiniteVolSym} we list the little groups for all possible total momentum assignments. In order to extract scattering information from lattice results, the quantization condition must also be projected to a fixed irrep, as we have done in eq.~\eqref{eq:QC}. This is discussed elsewhere in great detail; see for example refs.~\cite{\fvsg}. Here we only describe a few key points relevant to this work.

Restricting attention first to $\boldsymbol P = [000]$, and thus the $O_h$ group, and considering only the trivial irrep $A_{1g}$, our first aim is to work out the projectors introduced in eq.~\eqref{eq:QC}, denoted $\mathbb P_{A_{1g},\ell m}$ in this case. The defining property of $\mathbb P_{A_{1g},\ell m}$ is that, when the $m$ index is contracted with $Y_{\ell m}(\hat {\boldsymbol k})$, the resulting function is invariant under the group elements, i.e.~for any $R \in O_h$,
\begin{equation}
\label{eq:rotinvar}
\sum_m \mathbb P_{A_{1g},\ell m} Y_{\ell m}(\hat {\boldsymbol k}) = \sum_m \mathbb P_{A_{1g},\ell m} Y_{\ell m}(R \cdot \hat {\boldsymbol k}) \,,\end{equation}
where we stress that the sum only runs over $m$. One can inspect that the following quantity has this property
\begin{equation}
\label{eq:projdef}
\mathbb P_{A_{1g},\ell m} = \frac{1}{\mathcal N} \sum_{R \in O_h} Y^*_{\ell m}(R \cdot \hat {\boldsymbol e}) \,,
\end{equation}
where $\mathcal N$ is a normalization constant for nonzero vectors and $ \hat {\boldsymbol e}$ is a generic unit vector.
To see that the projector satisfies eq.~\eqref{eq:rotinvar}, note
\begin{align}
\sum_m \mathbb P_{A_{1g},\ell m} Y_{\ell m}(\hat {\boldsymbol k}) & = \sum_m \frac{1}{\mathcal N} \sum_{R' \in O_h} Y^*_{\ell m}(R' \cdot \hat {\boldsymbol e}) Y_{\ell m}(\hat {\boldsymbol k}) \,, \\
& = \frac{2 \ell+1}{4 \pi} \frac{1}{\mathcal N} \sum_{R' \in O_h} P_{\ell}(\hat {\boldsymbol k} \cdot R' \cdot \hat {\boldsymbol e}) \,, \\
& = \frac{2 \ell+1}{4 \pi} \frac{1}{\mathcal N} \sum_{R'' \in O_h} P_{\ell}(\hat {\boldsymbol k} \cdot R \cdot R'' \cdot \hat {\boldsymbol e}) \,, \\
& = \sum_m \mathbb P_{A_{1g},\ell m} Y_{\ell m}(R \cdot \hat {\boldsymbol k}) \,.
\end{align}
Similar relations can be used to prove
\begin{equation}
\label{eq:normconstant}
\vert \, \mathcal N \vert^2 = \vert O_h \vert \sum_{R \in O_h} \sum_m Y^*_{\ell m}(R \cdot \hat {\boldsymbol e}) Y_{\ell m}( \hat {\boldsymbol e}) \,.
\end{equation}

As an example, applying this for $\ell = 4$ gives
\begin{equation}
\mathbb P_{ A_{1g}, 4 m'} = \frac{1}{2 \sqrt{6}} \Big ( \sqrt{5} , 0 ,0 ,0, \sqrt{14}, 0 , 0, 0, \sqrt{5} \Big )_{m'} \,.
\end{equation}
This is then used to define $f_{40}$ and $f_{44}$, used in section~\ref{subsec:HPW}.

To make this useful for general $\ell$ we return to the expressions of section \ref{subsec:HPW}, beginning with the definition of $M_{\ell \ell'}$, eq.~\eqref{eq:Mdef}, which we repeat for convenience
\begin{align}
M^{\frak n}_{\ell \ell'} & = 4 \pi \mathbb P_{A_{1g},\ell m} \, \mathbb P_{A_{1g},\ell' m'} \frac{1}{\vert O_h \vert} \sum_{ R \in O_h } Y_{\ell m}(R \cdot \hat {\vec \nu}_{\frak n}) Y^*_{\ell' m'}(R \cdot \hat {\vec \nu}_{\frak n}) \,.
\end{align}
Now note that, because of eq.~\eqref{eq:rotinvar}, the sum over rotations is redundant and thus
\begin{align}
M^{\frak n}_{\ell \ell'} & = 4 \pi \mathbb P_{A_{1g},\ell m} \, \mathbb P_{A_{1g},\ell' m'} Y_{\ell m}( \hat {\vec \nu}_{\frak n}) Y^*_{\ell' m'}( \hat {\vec \nu}_{\frak n}) \,.
\end{align}
From this it is clear that $M^{\frak n}_{\ell \ell'}$ is rank one: We write $M^{\frak n}_{\ell \ell'} = \sqrt{\mathcal P_{\ell} } \sqrt{\mathcal P_{\ell'}}$ where
\begin{align}
\mathcal P_{\ell} & = 4 \pi \Big ( \mathbb P_{A_{1g},\ell m} Y_{\ell m}( \hat {\vec \nu}_{\frak n}) \Big )^2 \,.
\end{align}
Finally substituting our expression for the projector yields
\begin{align}
\mathcal P_{\ell} & = 4 \pi \bigg ( \frac{1}{\mathcal N} \sum_m \sum_{R \in O_h} Y^*_{\ell m}(R \cdot \hat {\boldsymbol e}) Y_{\ell m}( \hat {\vec \nu}_{\frak n}) \bigg )^2 \,, \\
& = 4 \pi \bigg ( \vert O_h \vert \sum_{R \in O_h} \sum_m Y^*_{\ell m}(R \cdot \hat {\boldsymbol e}) Y_{\ell m}( \hat {\boldsymbol e}) \bigg )^{-1} \bigg ( \sum_m \sum_{R \in O_h} Y^*_{\ell m}(R \cdot \hat {\boldsymbol e}) Y_{\ell m}( \hat {\vec \nu}_{\frak n}) \bigg )^2 \,.
\end{align}
where in the second line we have substituted the result for $\mathcal N$, eq.~\eqref{eq:normconstant}.

At this point, a few comments are in order about the arbitrary vector $ \hat {\boldsymbol e} $ in eq.~\eqref{eq:projdef}. To fully specify the role of this vector we take a small detour over some basics of groups and representations. While $\ell$ specifies an irrep of the group of continuous rotations $SO(3)$, it is no longer irreducible with respect to the subgroup of the finite volume system. Instead, a given $\ell$ value splits into a set of finite-volume irreps. This manifests in the quantization condition by the fact that finite-volume energies in the given irrep are shifted by the partial wave. For the case of $O_h$, the symmetry group of $\boldsymbol P =[000]$, all even $\ell$ besides $\ell=2$ contain at least one embedding of the trivial irrep. In addition, while exactly one embedding occurs for $\ell=0,4,6$ and $10$, for $\ell=12$ and some other higher values, multiple embeddings can arise.

Consider first the values of $\ell$ for which exactly one embedding appears. Then $ \mathbb P_{A_{1g},\ell m} $ must be uniquely specified up to a phase and any choice of $\hat {\boldsymbol e}$ will give the same result, up to that ambiguity. Within the definition of $\mathcal P_{\ell}$, it is particularly convenient to choose $\hat {\boldsymbol e} = \hat {\vec \nu}_{\frak n}$, from which follows
\begin{align}
\mathcal P_{\ell} & = 4 \pi \frac{1}{\vert O_h \vert} \sum_{R \in O_h} \sum_m Y^*_{\ell m}(R \cdot \hat {\boldsymbol k}) Y_{\ell m}( \hat {\vec \nu}_{\frak n}) \,, \\
& = (2 \ell+1) \frac{1}{\vert O_h \vert} \sum_{R \in O_h} P_{\ell}(\hat {\vec \nu}_{\frak n} \cdot R \cdot \hat {\vec \nu}_{\frak n}) \,.
\label{eq:PellFinal}
\end{align}
This matches eq.~\eqref{eq:Pellresult} of the main text.

Next consider a case such as $\ell=12$, for which multiple embeddings of the trivial irrep appear. Provided the state of interest does not exhibit an accidental degeneracy, one can show that using eq.~\eqref{eq:PellFinal} still gives the correct result for the leading-order shift. To prove this generally, suppose that a given $\ell$ value has $K > 1$ embeddings of $A_{1g}$ forming a basis of vectors $ \mathbb P_{A_{1g} (1),\ell m}, \mathbb P_{A_{1g} (2),\ell m}, \cdots, \mathbb P_{A_{1g} (K),\ell m}$. Choose $ \mathbb P_{A_{1g} (1),\ell m}$ as the vector generated by $\hat {\boldsymbol e} = \hat {\vec \nu}_{\frak n}$. Next note that, since the projectors are orthonormal, for any $k \neq 1$ one has
\begin{equation}
\sum_m \mathbb P_{A_{1g} (k),\ell m} \mathbb P^*_{A_{1g} (1),\ell m} = \frac{1}{\mathcal N} \sum_m \mathbb P_{A_{1g} (k),\ell m} \sum_{R \in O_h} Y_{\ell m}(R \cdot \hat {\vec \nu}_{\frak n}) = 0 \,.
\end{equation}
This implies that $\mathbb P_{A_{1g} (k),\ell m}$ annihilates the expression for $F_{\ell m, \ell' m'}$, expanded to leading-order about the non-accidentally degenerate state of interest, meaning that no solution appears for the given state and embedding. For states with an accidental degeneracy, by contrast, one must keep the full subspace spanned by $ \mathbb P_{A_{1g} (1),\ell m}, \mathbb P_{A_{1g} (2),\ell m}, \cdots, \mathbb P_{A_{1g} (K),\ell m}$ to identify the complete set of interacting solutions. For the rest frame this is a highly obscure case, since it requires $\ell=12$ or higher and the first accidentally degenerate state is the 8$^{\text{th}}$ excited state.

These results generalize readily to nonzero $\boldsymbol P$ with $A_{1g}$ replaced by $A_1$, the label of the trivial irrep for all groups besides $O_h$. As with the rest frame case, the key observation is that
\begin{equation}
\mathbb P_{A_{1},\ell m} = \frac{1}{\mathcal N} \sum_{R \in \LGP} Y^*_{\ell m}(R \cdot \hat {\boldsymbol \nu}_{\frak n}^\star) \,,
\end{equation}
defines a generic projector that can be used to derive the contribution of all partial waves to the leading-order energy shift.

\section{Equivalence of poles and non-interacting energies}
\label{app:DegenVecSolu}

In this appendix we show that the set of $\boldsymbol v$ solving
\begin{equation}
q^2 - \Gamma(\bfv|q, \bfd, L) = 0\,,
\end{equation}
exactly corresponds to the set satisfying
\begin{equation}
E = \omega_{\boldsymbol v} + \omega_{\boldsymbol d - \boldsymbol v} \,,
\end{equation}
provided that $E^\star > 0$.

As a first step we follow ref.~\cite{Kim:2005gf} to write
\begin{align}
q^2 - \Gamma(\bfv|q, \bfd, L) & = q^2 - \frac{1}{\gamma(q, \boldsymbol d, L)^2}\left(\bfv_\parallel - \frac{ \boldsymbol d }{2}\right)^2 - \bfv_\perp^2 \,, \\[5pt]
& = (E^\star/2 - \omega^\star_\bfv ) \times \Xi(\boldsymbol v, E, \boldsymbol d, L) \,,
\label{eq:qmGfinal}
\end{align}
where $\boldsymbol v_{\parallel}$ and $\boldsymbol v_\perp$ are the components of $\boldsymbol v$ that are parallel and perpendicular to $\boldsymbol d$, respectively. In the second line we have introduced
\begin{align}
\Xi(\boldsymbol v^\star, E, \boldsymbol d, L) & \equiv \frac{L^2}{4 \pi^2} \Big ( E^\star/2 + \omega^\star_\bfv \Big ) + 2 \frac{\boldsymbol d}{E} \cdot \boldsymbol v^\star - \frac{\boldsymbol d^2}{E^2} \left(E^\star/2- \omega^\star_\bfv \right) \,, \\[3pt]
& = \frac{L^2}{4\pi^2}\left[\frac{E^\star}{2}\left(1- \boldsymbol \beta^2\right)+\frac{4\pi}{L} \boldsymbol \beta \cdot \boldsymbol v^\star + \omega_{\boldsymbol v}^\star \big ( 1 + \boldsymbol \beta^2 \big ) \right] \,.
\end{align}
The result \eqref{eq:qmGfinal}, derived in ref.~\cite{Kim:2005gf}, together with the fact that $\Xi$ is finite for all finite values of its arguments (and nonzero $L$) is enough to show that a solution of $E^\star - 2 \omega_{\boldsymbol v}^\star=0$ also satisfies $q^2 - \Gamma = 0$.

To show the converse, that a solution of $q^2 - \Gamma = 0$ also satisfies $E^\star - 2 \omega_{\boldsymbol v}^\star=0$, we now prove that $\Xi$ cannot vanish for $E^\star >0$. This can be achieved by demonstrating the inequality
\begin{equation}
- 2 \boldsymbol \beta \cdot \boldsymbol k^\star - \omega_{\boldsymbol v}^\star \big ( 1 + \boldsymbol \beta^2 \big ) < \frac{E^\star}{2} \big ( 1 - \boldsymbol \beta^2 \big ) \,,
\end{equation}
where we have defined $\boldsymbol k^\star \equiv 2 \pi \boldsymbol v^\star/L$. This can be demonstrated via the equality
\begin{equation}
\label{eq:strongerineq}
- 2 \boldsymbol \beta \cdot \boldsymbol k^\star < \omega_{\boldsymbol v}^\star \big ( 1 + \boldsymbol \beta^2 \big ) \,,
\end{equation}
which is a stronger result since $1 - \boldsymbol \beta^2 > 0$.
To see that \eqref{eq:strongerineq} holds, note that the left-hand side is maximized when $- 2 \boldsymbol \beta \cdot \boldsymbol k^\star = 2 \vert \boldsymbol \beta \vert \vert \boldsymbol k^\star \vert$. Squaring both sides gives
\begin{equation}
4 \boldsymbol k^{\star 2} < (m^2 + \boldsymbol k^{\star 2} ) \frac{(1+ \boldsymbol \beta^2)^2}{ \boldsymbol \beta^2} \,,
\end{equation}
which holds since $x+1/x>2$ for $x \in [0,1)$. It follows that $\Xi \neq 0$ for all real values of its arguments, provided $E^\star > 0$.

Having shown that the set satisfying $q^2 - \Gamma = 0$ is equivalent to that satisfying $E^\star - 2 \omega_{\boldsymbol v}^\star = 0$, it remains only to prove that the latter is also equivalent to the set of $\boldsymbol v$ satisfying the moving frame condition: $E - \omega_{\boldsymbol v} - \omega_{\boldsymbol d - \boldsymbol v} = 0$. This is the case, due to the fact that the following expressions have the same set of roots for $E^\star > 0$
\begin{align}
E^\star - 2 \omega_{\boldsymbol v}^\star & \ \ \Leftrightarrow \ \ (E^\star - \omega_{\boldsymbol v}^\star + \omega_{\boldsymbol v}^\star) (E^\star - 2 \omega_{\boldsymbol v}^\star) \,, \\
& \ \ \Leftrightarrow \ \ (E^\star - \omega_{\boldsymbol v}^\star)^2 - \boldsymbol k^{\star 2} - m^2 \,, \\
& \ \ \Leftrightarrow \ \ (P-k)^2 - m^2 \,, \\
& \ \ \Leftrightarrow \ \ (E - \omega_{\boldsymbol v} - \omega_{\boldsymbol d - \boldsymbol v})(E - \omega_{\boldsymbol v} + \omega_{\boldsymbol d - \boldsymbol v}) \,,
\end{align}
where $\Leftrightarrow$ is used here to indicate that the two expressions have the same roots. Here we have introduced $P^{\star \mu} = (E^\star, \boldsymbol 0)$ and $k^{\star \mu} = (\omega_{\boldsymbol v}^\star, \boldsymbol k^\star)$ and used the fact that $(P-k)^2$ is a Lorentz scalar to rewrite it in the finite-volume frame.

Finally, to see that the unwanted factor in the final line does not induce any additional solutions, note that it has the same roots as the following:
\begin{align}
E - \omega_{\boldsymbol v} + \omega_{\boldsymbol d - \boldsymbol v} & \ \ \Leftrightarrow \ \ (E - \omega_{\boldsymbol v} + \omega_{\boldsymbol d - \boldsymbol v} ) (E + \omega_{\boldsymbol v} + \omega_{\boldsymbol d - \boldsymbol v} ) \\
& \ \ \Leftrightarrow \ \ (E+ \omega_{ \boldsymbol v'} )^2 - \omega_{\boldsymbol d - \boldsymbol v'}^2 \\
& \ \ \Leftrightarrow \ \ (P+k')^2 - m^2 \,, \\
& \ \ \Leftrightarrow \ \ (E^\star + \omega_{\boldsymbol k}'^\star)^2 - \omega_{\boldsymbol k}'^{\star 2} \,, \\
& \ \ \Leftrightarrow \ \ E^\star(E^\star +2 \omega_{\boldsymbol k}'^\star) \,,
\end{align}
where we have introduced $\boldsymbol v' = \boldsymbol d - \boldsymbol v$ and $k'^\mu = (\omega_{\boldsymbol k}', - 2 \pi \boldsymbol v' /L)$. The final expression is manifestly nonzero for $E^\star >0$. This completes the proof.

\section{Two observations concerning accidental degeneracy}
\label{app:ADobservations}

In this appendix, we prove the assertions given in section~\ref{subsec:AccDeg}.

We begin by demonstrating that, for total momentum types $\vec P = [00a]$ and $\vec P = [aaa]$, accidental degeneracies in the trivial irrep are only broken when $\ell = 4$ is included, even though $\ell=2$ contributes to the energies. We also show the converse, that for all other non-zero total momenta the degeneracy is broken by $\ell=2$.

Before turning to the various cases in the moving frame, we review the situation for $\vec P = [000]$. Here $\ell=4$ is the first non-trivial partial wave that contributes to trivial-irrep energies and, as we discuss in section~\ref{subsec:AccDeg}, including this partial wave does indeed split the accidentally degenerate states. The splitting occurs because the matrix $f_{\ell \ell'}$, truncated to $\ell_{\sf max}=4$ and then expanded to leading order about the non-interacting solution, has rank exceeding one whenever the solution of interest is accidentally degenerate.

The condition that the truncated and expanded $f_{\ell \ell'}$ is rank one is equivalent to the relation
\begin{align}
\label{eq:AccDegCondition}
\sum_{m = -\ell}^{\ell} \mathbb P_{A_{1g},\ell m} Y_{\ell m}(\hat \bfnu_\gothicn) = \sum_{m = -\ell}^{\ell} \mathbb P_{A_{1g},\ell m} Y_{\ell m}(\hat {\vec v}) \,, \qquad \forall \, \vec v \in \vec S_\gothicn \,,
\end{align}
and this holds for the low lying states, for which all elements of $\vec S_{\frak n}$ are rotations of $\bfnu_\gothicn$, but generally fails at $\ell=4$ whenever an accidental degeneracy occurs. For example, one can readily check that the left-hand side gives different values when evaluated at $\bfnu_\gothicn = (0,0,3)$ as compared to $\bfnu_\gothicn = (1,2,2)$. Since both of these three-vectors are within $\vec S_{\frak n}$ for the $8^{\text{th}}$ excited state, this is sufficient.

Returning to non-zero $\vec P$, note that $\boldsymbol \nu_{\frak n}$ and $\boldsymbol \nu'_{\frak n}$ represent degenerate states if and only if
\begin{equation}
\label{eq:EnergeyDeg}
\omega_{\boldsymbol \nu_{\frak n}} + \omega_{\boldsymbol d - \boldsymbol \nu_{\frak n}} = \omega_{\boldsymbol \nu'_{\frak n}} + \omega_{\boldsymbol d - \boldsymbol \nu'_{\frak n}} \,,
\end{equation}
but $\boldsymbol \nu_{\frak n}$ cannot be transformed into either $\boldsymbol \nu'_{\frak n}$ or $\boldsymbol d - \boldsymbol \nu'_{\frak n}$ via an element of the little group, $\LGP$.

Next, defining $\boldsymbol \nu_{\frak n}^\star$ as the result of boosting $(L \omega_{\boldsymbol \nu_{\frak n}}/(2 \pi), \boldsymbol \nu_{\frak n})$ with velocity $\boldsymbol \beta = - \boldsymbol P/(\omega_{\boldsymbol \nu_{\frak n}} + \omega_{\boldsymbol d - \boldsymbol \nu_{\frak n}})$, we see that eq.~\eqref{eq:EnergeyDeg} implies
\begin{equation}
2 \omega_{\boldsymbol \nu^\star_{\frak n}} =2 \omega_{\boldsymbol \nu'^\star_{\frak n}} \,,
\end{equation}
and thus\footnote{To avoid clutter of notation we drop the $\frak n$ subscript for the remainder of this appendix.}
\begin{equation}
(\boldsymbol \nu^\star_{\perp})^2 + (\boldsymbol \nu^\star_{\parallel})^2 = (\boldsymbol \nu'^\star_{\perp})^2 + (\boldsymbol \nu'^\star_{\parallel})^2 \,,
\end{equation}
where $\boldsymbol \nu^\star_{\perp}$ and $\boldsymbol \nu^\star_{\parallel}$ are three-vectors parallel and perpendicular to $\boldsymbol d$ satisfying $\boldsymbol \nu^\star = \boldsymbol \nu^\star_{\perp} + \boldsymbol \nu^\star_{\parallel}$. Because $\boldsymbol \nu^\star_{\parallel}$ and $ \boldsymbol \nu'^\star_{\parallel}$ are boost dependent, while $\boldsymbol \nu^\star_{\perp} = \boldsymbol \nu_{\perp} $ and $\boldsymbol \nu'^\star_{\perp} = \boldsymbol \nu'_{\perp} $ are not, one can additionally infer
\begin{equation}
\label{eq:ADequal}
(\boldsymbol \nu^\star_{\perp})^2 = (\boldsymbol \nu'^\star_{\perp})^2 \,, \qquad \qquad (\boldsymbol \nu^\star_{\parallel})^2 = (\boldsymbol \nu'^\star_{\parallel})^2 \,.
\end{equation}

We are now in position to determine whether the $\ell = 2$ partial wave will lead to a splitting in the corresponding energies. As with the rest-frame case, a splitting will occur whenever the $f_{\ell \ell'}$ matrix, truncated to $\ell_{\sf max} = 2$ and expanded about the state of interest, has a rank exceeding one. This, in turn, occurs whenever
\begin{align}
\label{eq:Y2mADcond}
\sum_{m = -2}^{2} \mathbb P_{A_{1},2 m} Y_{2 m}(\hat \bfnu^\star) \neq \sum_{m = -2}^{2} \mathbb P_{A_{1},2 m} Y_{2 m}(\hat \bfnu'^\star) \,,
\end{align}
where $\mathbb P_{A_{1},\ell m}$ is a projector to the trivial irrep of $\LGP$.

To explore this condition we require some additional notation. We define ${\nu}^\star_{\parallel, z } = \hat {\boldsymbol d} \cdot \boldsymbol \nu^\star_{\parallel}$ as the component of $\nu^\star_{\parallel}$ along the momentum direction. Note that this is equal to $\vert \boldsymbol \nu^\star_{\parallel} \vert$ up to a sign. Here we include the $z$-subscript to suggest the definition of a coordinate system with the $z$-axis along $\boldsymbol d$. Similarly we introduce $( {\nu}^\star_{\perp,x }, {\nu}^\star_{\perp,y } )$ as the two components of $\boldsymbol \nu^\star_{\perp}$ along two arbitrary axes perpendicular to $\boldsymbol d$. With these components in hand the sum over the $\ell = 2$ harmonics takes the form
\begin{multline}
\label{eq:Y2mexp}
\vert {\bfnu}^\star \vert^2 \sum_{m = -\ell}^{\ell} \mathbb P_{A_{1},\ell m} Y_{ \ell m}({\hat \bfnu}^\star ) = \alpha_1 \big [( {\nu}^\star_{\perp,x })^2 + ( {\nu}^\star_{\perp,y })^2 \big ] + \alpha_2 ( {\nu}^\star_{\parallel, z })^2 \\
+ \beta_1 \big [( {\nu}^\star_{\perp,x })^2 - ( {\nu}^\star_{\perp,y })^2 \big ]+ \beta_2 {\nu}^\star_{\perp,x } {\nu}^\star_{\perp,y } + \beta_3 {\nu}^\star_{\parallel, z } {\nu}^\star_{\perp,x } + \beta_4 {\nu}^\star_{\parallel, z } {\nu}^\star_{\perp,x } \,,
\end{multline}
where $\alpha_1$, $\alpha_2$, $\beta_1$, $\beta_2$, $\beta_3$, and $\beta_4$ are coefficients that depend on $\boldsymbol d$ and on the exact definition of $\mathbb P_{A_{1},2 m}$. Combining eqs.~\eqref{eq:ADequal}, \eqref{eq:Y2mADcond} and \eqref{eq:Y2mexp}, we deduce that the $\ell=2$ partial wave cannot break accidental degeneracies if, for a given $\boldsymbol P$, $\beta_i = 0$. Thus it remains only to show that this is the case for $\vec P = [00a]$ and $\vec P = [aaa]$ but not for other values of total momentum.

The terms multiplying $\beta_i$ coefficients are all symmetry breaking, i.e.~are not invariant under continuous rotations about the $\boldsymbol d$ axis. At the same time, because $\mathbb P_{A_{1},\ell m}$ projects to the trivial irrep, $\sum_m \mathbb P_{A_{1},\ell m} Y_{ \ell m}({\hat \bfnu}^\star ) $ must be invariant under the group elements, by construction. This gives a number of constraints on the coefficients, and it follows that $\beta_i = 0$ whenever the little group, $\LGP$, is large enough to give the required constraints. In the case of $\boldsymbol P = [00a]$ one can readily identify the required symmetries
\begin{gather}
{\nu}^\star_{\perp,x } \leftrightarrow {\nu}^\star_{\perp,y} \ \ \ \Rightarrow \ \ \ \beta_1 = 0 , \ \beta_3 = \beta_4 \,, \\
{\nu}^\star_{\perp,x } \leftrightarrow - {\nu}^\star_{\perp,x} \,, \ {\nu}^\star_{\perp,y } \leftrightarrow - {\nu}^\star_{\perp,y} \ \ \ \Rightarrow \ \ \ \beta_3 = \beta_4 = 0 \,, \\
{\nu}^\star_{\perp,x } \leftrightarrow - {\nu}^\star_{\perp,x} \ \ \ \Rightarrow \ \ \ \beta_2 = 0 \,.
\end{gather}
Similarly for $\boldsymbol P = [aaa]$ one can identify constraints from the six elements of the corresponding little group that ensure $\beta_i=0$. To complete the demonstration one must find examples of all other nonzero momentum types, namely types $[0aa]$, $[0ab]$, $[abb]$, $[abc]$, for which the $\beta_i$ coefficients are non-zero. We have done this through explicit calculation and have confirmed that whenever accidental degeneracy occurs, it is broken by the $\ell = 2$ harmonics. Note that this is highly plausible given the results of table~\ref{tab:FiniteVolSym}. Each of the symmetry groups for which $\ell = 2$ generates the splittings has 4 or fewer elements, meaning that not enough constraints arise to require $\beta_i = 0$.

We now turn to the second claim presented in section~\ref{subsec:AccDeg} that, in the limit where the coefficients of higher partial waves are taken arbitrarily small, one of the states in an accidentally degenerate system is shifted according to the naive result
\beq
\Delta^{(1)}_{E[\mathfrak{n}], +}(L) = g_{\gothicn} \frac{E^{(0)}_\gothicn (L) }{ 4 \omega_{\bfnu_\gothicn} \omega_{\vec d-\bfnu_\gothicn} }\frac{8 \pi a_0}{ \gamma_\mathfrak{n}^{(0)} L^3} \,.
\eeq
This follows directly from noting that the roots of eq.~\eqref{eq:TwoByTwoMatrix} (with the $\ell = 4$ replaced by a generic $\ell$) match those of the following relation:
\begin{equation}
\bigg [1 - f_{00}(q, \boldsymbol d, L) \frac{ \tan \delta_0(p)}{p} \bigg ] \bigg [1 - f_{\ell\ell}(q, \boldsymbol d, L) \frac{ \tan \delta_\ell(p)}{p} \bigg ] = f_{\ell0}(q, \boldsymbol d, L)^2 \frac{\tan \delta_0(p) \tan \delta_\ell(p)}{ p^2} \,.
\end{equation}
As $\delta_{\ell}(p)$ tends to zero this manifestly picks up the solution of eq.~\eqref{eq:SWaveMaster} together with non-interacting solutions.

\end{document}